\documentclass[a4paper,12pt]{article}

\pdfoutput=1
\usepackage{amsmath}
\usepackage{amssymb}
\usepackage{amsfonts}
\usepackage{mathrsfs}
\usepackage{bbm}
\usepackage{graphicx,subfigure,booktabs}
\usepackage[numbers,sort&compress]{natbib}
\usepackage{verbatim}
\usepackage{color}
\usepackage{ulem}
\usepackage{setspace}
\usepackage{multirow}
\usepackage{url}

\usepackage[utf8]{inputenc}

\usepackage[colorlinks,
linkcolor=black,
filecolor=black,
anchorcolor=black,
urlcolor=black,
citecolor=blue,
bookmarks=false,
]{hyperref}

\usepackage[hyphenbreaks]{breakurl}

\numberwithin{equation}{section}

\newlength{\dinwidth}
\newlength{\dinmargin}
\allowdisplaybreaks

\begin{document}

\title{\bf \Large \boldmath{$B_{s(d)}-\bar{B}_{s(d)}$} Mixing and \boldmath{$B_s\to\mu^+\mu^-$} Decay in the NMSSM with the Flavour Expansion Theorem}

\author{
Quan-Yi Hu\footnote{huquanyi@mail.ccnu.edu.cn},
Xin-Qiang Li\footnote{xqli@mail.ccnu.edu.cn},
Ya-Dong Yang\footnote{yangyd@mail.ccnu.edu.cn},\,
and
Min-Di Zheng\footnote{zhengmindi@mails.ccnu.edu.cn}\\[15pt]
\small Institute of Particle Physics and Key Laboratory of Quark and Lepton Physics~(MOE), \\[-2mm]
\small Central China Normal University, Wuhan, Hubei 430079, China}

\date{}
\maketitle
\vspace{-0.2cm}

\begin{abstract}
\noindent In this paper, motivated by the observation that the Standard Model predictions are now above the experimental data for the mass difference $\Delta M_{s(d)}$, we perform a detailed study of $B_{s(d)}-\bar{B}_{s(d)}$ mixing and $B_s\to\mu^+\mu^-$ decay in the $\mathbb{Z}_3$-invariant NMSSM with non-minimal flavour violation, using the recently developed procedure based on the Flavour Expansion Theorem, with which one can perform a purely algebraic mass-insertion expansion of an amplitude written in the mass eigenstate basis without performing any diagrammatic calculations in the interaction/flavour basis. Specifically, we consider the finite orders of mass insertions for neutralinos but the general orders for squarks and charginos, under two sets of assumptions for the squark flavour structures (\textit{i.e.}, while the flavour-conserving off-diagonal element $\delta_{33}^\text{LR}$ is kept in both of these two sectors, only the flavour-violating off-diagonal elements $\delta_{23}^\text{LL}$ and $\delta_{i3}^\text{RR}$ ($i=1,2$) are kept in the \text{LL} and \text{RR} sectors, respectively). Our analytic results are then expressed directly in terms of the initial Lagrangian parameters in the interaction/flavour basis, making it easy to impose the experimental bounds on them. It is found numerically that the NMSSM effects with the above two assumptions for the squark flavour structures can accommodate the observed deviation for $\Delta M_{s(d)}$, while complying with the experimental constraints from the branching ratios of $B_s\to \mu^+ \mu^-$ and $B\to X_s\gamma$ decays.
\end{abstract}

\newpage

\section{Introduction}
\label{sec:intro}

It is known that Supersymmetric~(SUSY) extensions of the Standard Model~(SM) are well motivated by being able to provide a unification of the SM gauge couplings at high scales, a solution of the hierarchy problem, and a viable dark matter candidate~\cite{Nilles:1983ge,Haber:1984rc}. As one of the low-energy realizations of SUSY, the Minimal Supersymmetric Standard Model (MSSM)~\cite{Dimopoulos:1981}, carrying the minimal field content consistent with observations, has received over years of continuous attentions; see \textit{e.g.} Refs.~\cite{Martin:1997ns,Chung:2003fi} for reviews. Despite having many advantages~\cite{Martin:1997ns,Chung:2003fi}, the MSSM needs to be extended because of the following two main motivations. The first one is due to the existence of the ``$\mu$-problem'' in the MSSM~\cite{Kim:1983dt}, where $\mu$ is a dimensionful parameter set by hand at the electroweak~(EW) scale before spontaneous symmetry breaking. The second one is driven by the recent discovered $125$~GeV SM-like Higgs boson~\cite{Aad:2012tfa,Chatrchyan:2012xdj}, which has imposed strong constrains on the parameter space of MSSM~\cite{Djouadi:2013vqa,Djouadi:2013uqa,Djouadi:2015jea}. Among the various non-minimal SUSY models, the Next-to-Minimal Supersymmetric Standard Model~(NMSSM)~\cite{Fayet:1975,Ellis:1989,Maniatis:2009re,Ellwanger:2009dp}, the simplest extension of the MSSM with a gauge singlet superfield, not only has the capability to fix these shortcomings of the MSSM, but also alleviates the tension implied by the lack of any evidence for superpartners below the EW scale~\cite{SUSY:search}. Specifically, this model can solve the ``$\mu$-problem'' of the MSSM elegantly by generating an effective $\mu$-term at the SUSY breaking scale~\cite{Barbieri:2006bg,Hall:2011aa}. Restrictions on the Higgs sector can also be relaxed in the NMSSM, because the Higgs field can acquire a larger tree-level mass with a low SUSY scale, making accordingly the quantum corrections to the observed $125$~GeV Higgs boson small~\cite{Maniatis:2009re,Ellwanger:2009dp}.

Along with the dedicated direct searches at the Large Hadron Collider~(LHC) for SUSY particles~\cite{SUSY:search}, it is also interesting and even complementary to investigate the virtual effects of these hypothesized particles on the low-energy processes, such as the neutral-meson mixings as well as the CP violation and rare decays of various hadrons~\cite{Buras:2000dm,Buras:2001mb,Buras:2001ra,Buras:2002vd,Buras:2002wq,Buras:2003td,Hiller:2004ii,Domingo:2007dx,Altmannshofer:2007cs,Hodgkinson:2008qk,AranaCatania:2011ak,Arana-Catania:2014ooa,Crivellin:2015dta,Domingo:2015wyn,Blanke:2016bhf,Domingo:2018qfg}. In this paper, we shall focus on the $\mathbb{Z}_3$-invariant NMSSM~\cite{Maniatis:2009re,Ellwanger:2009dp}, a simpler scenario of NMSSM with a scale-invariant superpotential, and study its effects on $B_{s(d)}-\bar{B}_{s(d)}$ mixing, $B_s\to\mu^+\mu^-$ and $B\to X_s\gamma$ decays.

The strength of $B_{s(d)}-\bar{B}_{s(d)}$ mixing is described by the mass difference $\Delta M_{s(d)}$ between the two mass eigenstates of neutral $B_{s(d)}$ mesons, and the uncertainties of the bag parameters and decay constants, which quantify the hadronic matrix elements of local four-quark operators between $|B_{s(d)}\rangle$ and $|\bar{B}_{s(d)}\rangle$ states, make up the largest uncertainty by far in the prediction of $\Delta M_{s(d)}$~\cite{Artuso:2015swg,Jubb:2016mvq}. These parameters can be determined either by lattice simulations~\cite{Dalgic:2006gp,Gamiz:2009ku,Bouchard:2011xj,Carrasco:2013zta,Dowdall:2014qka,Bazavov:2016nty,Boyle:2018knm} or with Heavy Quark Effective Theory (HQET) sum rules~\cite{Grozin:2016uqy,Grozin:2017uto,Kirk:2017juj,Grozin:2018wtg,King:2019lal}, with their results being now compatible in precision with each other~\cite{Aoki:2019cca,King:2019lal}. Depending on the values of these input parameters as well as the Cabibbo-Kobayashi-Maskawa (CKM) matrix elements~\cite{Cabibbo:1963yz,Kobayashi:1973fv}, the SM prediction of $\Delta M_{s(d)}$ varies from being consistent with to being larger than the experimental data; see \textit{e.g.} Refs.~\cite{Blanke:2016bhf,DiLuzio:2017fdq,Blanke:2018cya} for detailed discussions. In particular, taking the latest lattice averages for these non-perturtive parameters from the Flavour Lattice Averaging Group (FLAG)~\cite{Aoki:2019cca}, which are dominated by the values of Fermilab Lattice and MILC (FNAL/MILC) collaboration in 2016~\cite{Bazavov:2016nty}, one gets~\cite{DiLuzio:2017fdq,Bazavov:2016nty}\footnote{While the 2016 FNAL/MILC derived values of the bag parameters still need an independent cross-check by other lattice groups or by other methods, we shall adopt in this paper its direct results for the matrix elements, \textit{i.e.} $f_{B_{s(d)}}\sqrt{B_{s(d)}}$, which profit from the full set of correlations among them and are, therefore, more suitable for phenomenological applications~\cite{Bazavov:2016nty}.}
\begin{equation}\label{eq:smDMq}
\Delta M^\text{SM}_s = 20.01 \pm 1.25 \, \text{ps}^{-1} ,
\quad
\Delta M^\text{SM}_d = 0.630 \pm 0.069 \, \text{ps}^{-1} ,
\end{equation}  
which are about $1.8~\sigma$ above their respective experimental values~\cite{Tanabashi:2018oca,Amhis:2016xyh}
\begin{equation}\label{eq:expDMq}
\Delta M^\text{exp}_s = 17.757 \pm 0.021 \, \text{ps}^{-1} ,
\quad
\Delta M^\text{exp}_d = 0.5064 \pm 0.0019 \, \text{ps}^{-1}.
\end{equation}
This observation has profound implications for New Physics~(NP) models~\cite{Blanke:2016bhf,DiLuzio:2017fdq,Blanke:2018cya}. As detailed in Refs.~\cite{Altmannshofer:2007cs,Blanke:2016bhf,DiLuzio:2017fdq,Blanke:2018cya}, this implies particularly that the constrained minimal flavour violating~(CMFV) models, in which all flavour violations arise only from the CKM matrix, have difficulties in describing the current data on $\Delta M_{s(d)}$. Thus, in order to reconcile such a tension, one has to resort to the scenarios with non-minimal flavour violation~(NMFV), which involve extra sources of flavour- and/or CP-violation and can, therefore, provide potential negative contributions to $\Delta M_{s(d)}$~\cite{Foster:2005wb,Kumar:2016vhm,Ghosh:2015nui}. This motivates us to investigate whether the $\mathbb{Z}_3$-invariant NMSSM with NMFV, in which the extra flavour violations arise from the non-diagonal parts of the squark mass matrices related to the soft SUSY breaking terms, can accommodate the observed deviation for $\Delta M_{s(d)}$.

In SUSY models, a physical transition amplitude is more conveniently calculated in the interaction/flavour basis in which all gauge interactions are flavour diagonal and the flavour-changing interactions originate from the off-diagonal entries of the mass matrices in the initial Lagrangian before diagonalization and identification of the physical states, than in the mass eigenstate (ME) basis in which the amplitude is expressed in terms of the physical masses and mixing matrices. This can be achieved with the following two different methods. The first one is based on the well-known diagrammatic technique called the Mass Insertion Approximation~(MIA)~\cite{Hall:1985dx,Hagelin:1992tc,Gabbiani:1996hi,Misiak:1997ei}. Here the diagonal elements of the mass matrices are absorbed into the definition of the (un-physical) massive propagators and the amplitude is, at each loop order, expanded into an infinite series of the off-diagonal elements of the mass matrices, commonly referred to as the mass-insertion~(MI) parameters. The second one is based on the Flavour Expansion Theorem~(FET)~\cite{Dedes:2015twa}, according to which an analytic function about zero of a Hermitian matrix can be expanded polynomially in terms of its off-diagonal elements with coefficients being the divided differences of the analytic function and arguments the diagonal elements of the Hermitian matrix. Being a purely algebraic method, it offers an alternative derivation of the MIA result directly from the amplitude calculated in the ME basis, without performing the tedious and error-prone diagrammatic calculations with MIs in the interaction/flavour basis~\cite{Dedes:2015twa}. Even in the case where there is no clear diagrammatic picture, the FET expansion can still give a consistent MIA result. This method has also been automatized in the package {\tt MassToMI}~\cite{Rosiek:2015jua}, facilitating the expansion of an ME amplitude to any user-defined MI order. See \textit{e.g.} Refs.~\cite{Dedes:2015twa,Dedes:2014asa,Eberl:2016aox,Kumar:2016vhm,Crivellin:2018mqz} for recent applications of this method.

In the $\mathbb{Z}_3$-invariant NMSSM with NMFV, we further assume that the direct mixing between the first two generations of squarks is absent in the mass squared matrices\footnote{Note that, even in such a case, the simultaneous presence of non-zero $1 \text{-} 3$ and $2 \text{-} 3$ mixings can induce the $1 \text{-} 2$ mixing starting at the second order in the MI expansion. This effect is, however, too small to violate the constraint from Kaon physics~\cite{Arana-Catania:2014ooa,Kowalska:2014opa,DeCausmaecker:2015yca,Jager:2008fc}.}, which is strongly suppressed by the data on $K^0-\bar{K}^0$ mixing and rare kaon decays~\cite{Arana-Catania:2014ooa,Kowalska:2014opa,DeCausmaecker:2015yca,Jager:2008fc}. For example, under the constraint from $K^0-\bar{K}^0$ mixing and even with a rough estimate of the hadronic matrix elements, the MI parameters of the first two generations are bounded to be smaller than $0.04$ in both the \text{LL} and \text{RR} sectors~\cite{Jager:2008fc,Misiak:1997ei,Gabbiani:1996hi}. As we focus only on the flavour mixing effects observed in $B_{s(d)}-\bar{B}_{s(d)}$ mixing as well as $B_s\to\mu^+\mu^-$ and $B\to X_s\gamma$ decays~\cite{Buras:2002vd,Buras:2002wq,Buras:2003td,Hiller:2004ii,Domingo:2007dx,Altmannshofer:2007cs,Hodgkinson:2008qk,AranaCatania:2011ak,Arana-Catania:2014ooa,Crivellin:2015dta,Domingo:2015wyn,Blanke:2016bhf,Domingo:2018qfg}, it is natural to assume that the third-generation squarks can mix with the other two generations simultaneously. With such specific squark flavour structures, we shall then adopt the FET procedure~\cite{Dedes:2015twa} to calculate the mass difference $\Delta M_{s(d)}$ and the branching ratio of $B_s\to \mu^+ \mu^-$ decay. It should be noted that the mass matrices of squarks~(see Eqs.~\eqref{eq:MUi}-\eqref{eq:MDii}) and charginos~(see Eq.~\eqref{eq:Mchi}) can be diagonalized analytically, and only the neutralino mass matrix~(see Eq.~\eqref{eq:Mchi0}), being a $5\times 5$ matrix, has to be done numerically. For the squarks and charginos sectors, both the full FET approach and the calculation directly in the ME basis can be used to get an analytical result of the observables considered. But for the neutralino sector, an analytical result can be obtained only in the FET approach. As a purely algebraic method, the FET approach allows for a systematic expansion of a transition amplitude in terms of the MI parameters. Thus, the result obtained is a polynomial with the MI parameters as the variables, and hence allows for a more transparent understanding of the qualitative behaviour of the transition amplitude. In addition, being expressed directly in terms of the initial Lagrangian parameters in the interaction/flavour basis, they can be conveniently and easily exploited to put the experimental bounds on the parameters of the initial Lagrangian in this way. These features are, however, lost in the result obtained directly in the ME basis, in which the MI parameters are usually contained in the denominators and the arguments of various logarithms. On the other hand, one should note that the FET results agree very well with the ones calculated directly in the ME basis, as will be demonstrated in Sec.~\ref{sssec:FETcheck}. So, for consistency, we shall adopt the FET approach throughout this paper. Specifically, we consider the general MI orders for squarks and charginos but the finite MI orders for neutralinos. Although the general MI orders have been considered in Refs.~\cite{Arganda:2017vdb,Herrero:2018luu}, only one kind of MI parameter is kept in the whole ``fat propagators''. In our case, however, there exist two kinds of MI parameters in each line and the mixed arrangement of them is required. For concreteness, we call our procedure the FET expansion with different MI order and, by checking if the FET results agree with the ones calculated directly in the ME basis, test our estimation for the optimal cutting-off MI orders. For the branching ratio of $B\to X_s\gamma$ decay, the public code {\tt SUSY\_FLAVOR}~\cite{Rosiek:2010ug,Crivellin:2012jv,Rosiek:2014sia} is used.

Our paper is organized as follows. In Sec.~\ref{sec:2}, after specifying the flavour structures assumed throughout this paper, we introduce the FET procedure with different MI order, which is then used to calculate the $B_{s(d)}-\bar{B}_{s(d)}$ mixing and $B_s\to\mu^+\mu^-$ decay in Sec.~\ref{sec:3}. Detailed numerical results and discussions are then presented in Sec.~\ref{sec:4}. Our conclusions are finally made in Sec.~\ref{sec:conclusions}. For convenience, the block terms of squarks and charginos are listed in the appendix.

\section{\boldmath{$\mathbb{Z}_3$}-invariant NMSSM and the FET procedure}\label{sec:2}

\subsection{Lagrangian of the \boldmath{$\mathbb{Z}_3$}-invariant NMSSM}\label{sec:Z3-invariant NMSSM}

At the Lagrangian level, the $\mathbb{Z}_3$-invariant NMSSM differs from the MSSM by the superpotential and the soft SUSY breaking part. The scale-invariant superpotential of NMSSM reads~\cite{Baglio:2013iia,Allanach:2008qq}
\begin{align}\label{eq:WNMSSM}
W_\text{NMSSM} = W_\text{MSSM}\big|_{\mu=0} + \lambda\,\hat{S}\,\hat{H}_u \cdot \hat{H}_d + \frac{1}{3}\,\kappa\,\hat{S}^3,
\end{align}
where $W_\text{MSSM}\big|_{\mu=0}$ is the MSSM superpotential but without the $\mu$ term~\cite{Rosiek:1990kg,Rosiek:1995kg}, and $\hat{S}$ denotes the Higgs singlet superfield, while $\hat{H}_u=(\hat{H}_u^+,\hat{H}_u^0)^T$ and $\hat{H}_d=(\hat{H}_d^0,\hat{H}_d^-)^T$ are the two Higgs doublet superfields, with the convention $\hat{H}_u \cdot \hat{H}_d=\hat{H}_u^+\hat{H}_d^- - \hat{H}_u^0 \hat{H}_d^0$. The dimensionless parameters $\lambda$ and $\kappa$ can be complex in general, but are real in the CP-conserving case. After the scalar component of $\hat{S}$ gets a non-zero vacuum expectation value~(VEV), $\langle S\rangle=v_s/\sqrt{2}$, the second term in Eq.~\eqref{eq:WNMSSM} generates an effective $\mu$ term, with $\mu_\text{eff}=\lambda v_s/\sqrt{2}$, which then solves the ``$\mu$-problem'' of the MSSM~\cite{Maniatis:2009re,Ellwanger:2009dp}.

With the scalar components of the Higgs doublet and singlet superfields being denoted by $H_u$, $H_d$, and $S$, respectively, the soft SUSY breaking Lagrangian of the $\mathbb{Z}_3$-invariant NMSSM is then given by~\cite{Baglio:2013iia,Allanach:2008qq}
\begin{align}\label{eq:Lsoft}
-\mathcal{L}^\text{NMSSM}_\text{soft} = -\mathcal{L}^\text{MSSM}_\text{soft}\big|_{\mu=0} + m^2_S\,|S|^2 + \left(\lambda\,A_\lambda\,S\,H_u \cdot H_d + \frac{1}{3}\,\kappa\,A_\kappa\,S^3 + \text{h.c.}\right),
\end{align}
where $\mathcal{L}^\text{MSSM}_\text{soft}\big|_{\mu=0}$ corresponds to the MSSM part but with the $\mu$-related term removed~\cite{Rosiek:1990kg,Rosiek:1995kg}. While the mass parameter $m^2_S$ is real, the trilinear couplings $A_\lambda$ and $A_\kappa$ are generally complex, but are also real in the CP-conserving case, as is assumed throughout this paper.

\subsection{Flavour structures of the \boldmath{$\mathbb{Z}_3$}-invariant NMSSM}\label{sec:FlavorStructure}

Firstly, we focus on the up- and down-squark mass squared matrices $M_{\tilde{U}}^2$ and $M_{\tilde{D}}^2$, which can be written in their most general $2\times 2$-block form as~\cite{Allanach:2008qq}
\begin{align}
M_{\tilde{q}}^2=\begin{pmatrix}
M_{\tilde{q},\text{LL}}^2 & M_{\tilde{q},\text{LR}}^2\\
M_{\tilde{q},\text{RL}}^2 & M_{\tilde{q},\text{RR}}^2
\end{pmatrix},\qquad \tilde{q}=\tilde{U},\tilde{D},
\end{align}
in the so-called super-CKM basis~\cite{Misiak:1997ei}. In the NMFV paradigm~\cite{AranaCatania:2011ak,Arana-Catania:2014ooa}, these two mass matrices are not yet diagonal and can introduce general squark flavour mixings that are usually described by a set of dimensionless parameters $\delta_{ij}^\text{AB}$, with $\text{A,\,B=L,\,R}$ referring to the left- and right-handed superpartners of the corresponding quarks and $i,\,j=1,\,2,\,3$ the generation indices~\cite{Gabbiani:1996hi}.

Throughout this paper, we assume that the third-generation squarks can mix with the other two generations simultaneously, while the direct mixing between the latter two, which is strongly suppressed by the data on Kaon physics~\cite{Arana-Catania:2014ooa,Kowalska:2014opa,DeCausmaecker:2015yca,Jager:2008fc}, is absent in the squark mass squared matrices. It is also noted that the off-diagonal elements in the \text{LR} and \text{RL} sectors are severely restricted by the dangerous charge and colour breaking (CCB) minima and unbounded from below directions (UFB) in the effective potential~\cite{Casas:1995pd,Casas:1996de,Casas:1997ze,Ellwanger:1999bv}. Here we set all the flavour-violating off-diagonal elements in the \text{LR} and \text{RL} sectors to be zero, because the CCB and UFB bounds on them are always stronger than the ones from flavour-changing neutral-current processes, and even do not decrease when the SUSY scale increases~\cite{Casas:1995pd,Casas:1996de,Casas:1997ze}. These upper bounds on the flavour-conserving off-diagonal elements are, however, weaker than on the flavour-violating counterparts, which are both proportional to the related running quark masses~\cite{Ellwanger:1999bv,Bartl:2014bka}. So we only keep $\delta_{33}^\text{LR}$ non-zero and set all the other flavour-conserving off-diagonal elements to be zero, as the upper bounds on the latter are much stronger than on $\delta_{33}^\text{LR}$. In addition, a non-zero $\delta_{33}^\text{LR}$ is needed to reproduce the observed $125$~GeV SM-like Higgs boson~\cite{Arana-Catania:2014ooa,Kowalska:2014opa,DeCausmaecker:2015yca}. All the above observations promote us to consider the following two sets of squark flavour structures: while the flavour-conserving off-diagonal element $\delta_{33}^\text{LR}$ is kept in both of these two sectors, only the flavour-violating off-diagonal elements $\delta_{23}^\text{LL}$ and $\delta_{i3}^\text{RR}$~($i=1,2$) are kept in the \text{LL} and \text{RR} sectors, respectively. Then, the two mass squared matrices $M_{\tilde{U}}^2$ and $M_{\tilde{D}}^2$ in case I, in which the flavour violation resides only in the \text{LL} sector, are given, respectively, by
\begin{align}
M_{\tilde{U}}^2&=\begin{pmatrix}
{M_S}_1 & 0 & 0  & 0 & 0 & 0\\
0 &  {M_S}_1 & \delta_{23} \sqrt{{M_S}_1 {M_S}_2} & 0 & 0 & 0\\
0 &  \delta_{23} \sqrt{{M_S}_1 {M_S}_2} & {M_S}_2  & 0 & 0 & \delta_{36} {M_S}_2\\
0 & 0 & 0 & {M_S}_1 & 0 & 0\\
0 & 0 & 0 & 0 & {M_S}_1 & 0\\
0 & 0 & \delta_{36} {M_S}_2 & 0 & 0 & {M_S}_2\\
\end{pmatrix},\label{eq:MUi}\\[3mm]
M_{\tilde{D}}^2&=\begin{pmatrix}
{M_S}_1 & 0 & -\lambda_\text{CKM}\delta_{23} \sqrt{{M_S}_1 {M_S}_2}  & 0 & 0 & 0\\
0 &  {M_S}_1 & \delta_{23} \sqrt{{M_S}_1 {M_S}_2} & 0 & 0 & 0\\
-\lambda_\text{CKM}\delta_{23} \sqrt{{M_S}_1 {M_S}_2} &  \delta_{23} \sqrt{{M_S}_1 {M_S}_2} & {M_S}_2 & 0 & 0 & 0\\
0 & 0 & 0 & {M_S}_1 & 0 & 0\\
0 & 0 & 0 & 0 & {M_S}_1 & 0\\
0 & 0 & 0 & 0 & 0 & {M_S}_2\\
\end{pmatrix},\label{eq:MDi}
\end{align}
where $\delta_{23}\equiv\delta_{23}^\text{LL}$ and $\delta_{36}\equiv \delta_{33}^\text{LR}$. In the \text{LL} sectors, which satisfy the relation $(M_{\tilde{D}}^2)_\text{LL} = K^\dagger (M_{\tilde{U}}^2)_\text{LL} K$~(with $K$ being the CKM matrix) due to the $SU(2)_L$ gauge invariance~\cite{Misiak:1997ei}, we have neglected safely the ${\cal O}(\lambda_\text{CKM}^2)$ terms in Eq.~\eqref{eq:MDi} (and also in Eq.~\eqref{eq:MDii}), where $\lambda_{\text{CKM}}=|V_{us}|$ is the expansion parameter in the Wolfenstein parameterization~\cite{Wolfenstein:1983yz} of the CKM matrix. Here we have also assumed that the first two generations of squarks are nearly degenerate in mass~\cite{Martin:1997ns}.

In case II with the flavour violation arising only from the \text{RR} sector, on the other hand, the two mass squared matrices are given, respectively, by
\begin{align}
M_{\tilde{U}}^2&=\begin{pmatrix}
{M_S}_1 & 0 & 0  & 0 & 0 & 0\\
0 & {M_S}_1 & 0 & 0 & 0 & 0\\
0 & 0 & {M_S}_2 & 0 & 0 & \delta_{36} {M_S}_2\\
0 & 0 & 0 & {M_S}_1 & 0 & 0   \\
0 & 0 & 0 & 0 & {M_S}_1 & 0   \\
0 & 0 & \delta_{36} {M_S}_2 & 0 & 0 & {M_S}_2 \\
\end{pmatrix},\label{eq:MUii}\\[3mm]
M_{\tilde{D}}^2&=\begin{pmatrix}
{M_S}_1 & 0 & 0  & 0 & 0 & 0\\
0 &  {M_S}_1 & 0 & 0 & 0 & 0\\
0 &  0 & {M_S}_2 & 0 & 0 & 0\\
0 & 0 & 0 & {M_S}_1 & 0 & \delta_{46} \sqrt{{M_S}_1 {M_S}_2}\\
0 & 0 & 0 & 0 & {M_S}_1 & \delta_{56} \sqrt{{M_S}_1 {M_S}_2}\\
0 & 0 & 0 & \delta_{46} \sqrt{{M_S}_1 {M_S}_2} & \delta_{56} \sqrt{{M_S}_1 {M_S}_2} & {M_S}_2\\
\end{pmatrix},\label{eq:MDii}
\end{align}
where $\delta_{46}\equiv \delta_{13}^{\text{RR}}$ and $\delta_{56}\equiv \delta_{23}^{\text{RR}}$. Here, for simplicity, we have assumed that all the $\delta_{ij}^\text{AB}$ parameters are real and hence $\delta_{ij}^\text{AB}=\delta_{ji}^\text{BA}$, due to the hermiticity of the squark mass matrices.

The mass matrix for charginos in the interaction basis reads~\cite{Rosiek:1995kg}
\begin{align}\label{eq:Mchi}
M_{\chi}=\begin{pmatrix}
M_2 & \sqrt{2} m_W \sin \beta\\
\sqrt{2} m_W \cos \beta & \mu_\text{eff}
\end{pmatrix},
\end{align}
where $M_2$ is the wino mass, and $\beta=\tan^{-1}(v_u/v_d)$ is the mixing angle of the two Higgs doublets, defined in terms of their VEVs $v_u=\sqrt2\,\langle H_u\rangle$ and $v_d=\sqrt2\,\langle H_d\rangle$. The squared masses ${M_C}_i$, ${M_P}_i$ and the MI parameters $\delta^C_{ij}$, $\delta^P_{ij}$ are defined, respectively, by
\begin{align}\label{eq:Mchidiag}
{M_C}_i & = (M_\chi^\dag M_\chi)_{ii} ,
&
{M_P}_i & =(M_\chi M_\chi^\dag)_{ii} ,
\\[2mm]
\delta^C_{ij} & = \frac{(M_\chi^\dag M_\chi)_{ij}}{\sqrt{{M_C}_i {M_C}_j}} ,
&
\delta^P_{ij} & = \frac{(M_\chi M_\chi^\dag)_{ij}}{\sqrt{{M_P}_i {M_P}_j}} ,\label{eq:Mchinondiag}
\end{align}
where $i\neq j$ and the summation is not applied for the same indices here.

The neutralino mass matrix is given in the basis $(\tilde{B}$, $\tilde{W}^3$, $\tilde{H}_d^0$, $\tilde{H}_u^0$, $\tilde{S})^T$ by~\cite{Kumar:2016vhm,Maniatis:2009re}
\begin{equation}\label{eq:Mchi0}
M_{\chi_0}=\left( \begin{array}{ccccc}
M_1 &0 &-\frac{e v_d}{2 \cos\theta_W} & \frac{e v_u}{{2} \cos\theta_W}  & 0  \\
0 &M_2   & \frac{e v_d}{2 \sin\theta_W} & -\frac{e v_u}{2 \sin\theta_W} &0 \\
-\frac{e v_d}{2 \cos\theta_W} & \frac{e v_d}{2 \sin\theta_W} & 0 & - \mu_\text{eff} &  -\frac{\lambda v_u}{\sqrt{2}}\\
\frac{e v_u}{{2} \cos\theta_W} & -\frac{e v_u}{2 \sin\theta_W} & - \mu_\text{eff} & 0 &  - \frac{\lambda v_d}{\sqrt{2}} \\
0 & 0 & -\frac{\lambda v_u}{\sqrt{2}} & - \frac{\lambda v_d}{\sqrt{2}} &  {\sqrt 2} \kappa v_s \end{array} \right),
\end{equation}
where $M_1$ is the bino mass, and $\theta_W$ is the weak mixing angle. Such a mass matrix indicates that the singlino $\tilde{S}$ couples only to the Higgsinos $\tilde{H}_d^0$ and $\tilde{H}_u^0$, but not to the gauginos $\tilde{B}$ and $\tilde{W}^3$~\cite{Maniatis:2009re}. The squared masses ${M_N}_i$ and the MI parameters $\delta^N_{ij}$ are defined, respectively, by
\begin{align}\label{eq:Mchi0diag}
{M_N}_i = (M_{\chi_0}^\dag M_{\chi_0})_{ii} , \qquad
\delta^N_{ij} = \frac{(M_{\chi_0}^\dag M_{\chi_0})_{ij}}{\sqrt{{M_N}_{i} {M_N}_{j}}} ,
\end{align}
where $i\neq j$ and the summation is also not applied for the same indices. Diagonalization of Eq.~\eqref{eq:Mchi0} is rather involved and has to be in practice performed numerically~\cite{Maniatis:2009re}.

\subsection{FET expansion with different MI order}\label{subsec:HMIFETpro}

Before performing the FET expansion, one has to write down the transition amplitude in the ME basis~\cite{Dedes:2015twa}. All the relevant Feynman rules are taken from Refs.~\cite{Rosiek:1995kg,Ellwanger:2009dp,Maniatis:2009re,Franke:1995tc,Ellwanger:2004xm}. Then, the procedure of FET expansion with general/finite MI order includes the following three steps.

\textbf{Step 1}. One transforms the amplitude written in the ME basis into the intermediate result expressed in terms of the blocks $L_\text{X}(i,j)$, which are defined, respectively, as~\cite{Rosiek:2015jua}
\begin{align}\label{eq:blockQ}
\sum_A (Z_U)_{iA} l_p(m_A^2) (Z_U)^*_{jA} & = l_p(M_{\tilde{U}}^2)_{ij} \equiv L_\text{U}(i,j),\\[2mm]
\sum_A (Z_D)_{iA} l_p(m_A^2) (Z_D)^*_{jA} & = l_p(M_{\tilde{D}}^2)_{ij} \equiv L_\text{D}(i,j),
\end{align}
for the up and down squarks behaving as scalar fields,
\begin{align}\label{eq:blockchargino}
\sum_A (Z_{\chi}^+)_{iA} l_p(m_A^2) (Z_{\chi}^+)^*_{jA} & = l_p(M_\chi^{\dag} M_\chi)_{ij} \equiv L_\text{C}(i,j) ,\\[2mm]
\sum_A (Z_{\chi}^-)_{iA} l_p(m_A^2) (Z_{\chi}^-)^*_{jA} & = l_p(M_\chi M_\chi^{\dag})_{ij} \equiv L_\text{P}(i,j) ,\\[2mm]
\sum_A (Z_{\chi}^-)_{iA} m_A l_p(m_A^2) (Z_{\chi}^+)^*_{jA} & = \sum_k (M_\chi)_{ik}\, l_p(M_\chi^{\dag} M_\chi)_{kj} \equiv \sum_k (M_\chi)_{ik}\, L_\text{C}(k,j) ,\\
\sum_A (Z_{\chi}^+)_{iA} m_A l_p(m_A^2) (Z_{\chi}^-)^*_{jA} & = \sum_k (M_\chi^{\dag})_{ik}\, l_p(M_\chi M_\chi^{\dag})_{kj} \equiv \sum_k (M_\chi^{\dag})_{ik}\, L_\text{P}(k,j),
\end{align}
for the charginos that behave as Dirac fermions, and
\begin{align}
\sum_A (Z_{\chi_0})_{iA} l_p(m_A^2) (Z_{\chi_0})^*_{jA} & = l_p(M_{\chi_0}^{\dag} M_{\chi_0})_{ij} \equiv L_\text{N}(i,j) ,\\
\sum_A (Z_{\chi_0})_{iA} m_A l_p(m_A^2) (Z_{\chi_0})_{jA} & = \sum_k (M_{\chi_0}^{\dag})_{ik}\, l_p(M_{\chi_0}^{\dag} M_{\chi_0})_{kj} \equiv \sum_k (M_{\chi_0}^{\dag})_{ik}\, L_\text{N}(k,j) ,\\
\sum_A (Z_{\chi_0})^*_{iA} m_A l_p(m_A^2) (Z_{\chi_0})^*_{jA} & = \sum_k (M_{\chi_0})_{ik}\, l_p(M_{\chi_0}^{\dag} M_{\chi_0})_{kj} \equiv \sum_k (M_{\chi_0})_{ik}\, L_\text{N}(k,j),\label{eq:blockneutralino}
\end{align}
for the neutralinos behaving as Majorana fermions. Here $l_p(m_A^2)$ represents symbolically part of the transition amplitude that depends on the mass $m_A$ of an internal physical particle in a Feynman diagram, at tree or loop level; for example, $l_p(m_A^2)=1/(q^2-m_A^2)$ can be a propagator, with $q^\mu$ being the momentum of the particle $A$. The unitary transformation matrices $Z_U$, $Z_D$, $Z_{\chi}^+$, $Z_{\chi}^-$, and $Z_{\chi_0}$ are introduced to diagonalize the Hermitian mass squared matrices $M_{\tilde{U}}^2$, $M_{\tilde{D}}^2$, $M_\chi^\dag M_\chi$, $M_\chi M_\chi^\dag$, and $M_{\chi_0}^\dag M_{\chi_0}$, respectively. We use $A,B,\cdots$ and $i,j,k,\cdots$ to represent the flavour indices of field multiplets in the ME and the interaction/flavour basis, respectively.

After applying the transformation rules specified by Eqs.~\eqref{eq:blockQ}--\eqref{eq:blockneutralino}, one can see that the blocks $L_\text{X}(i,j)$ depend only on the matrix elements of some functions with arguments being the Hermitian mass squared matrices, and can be expanded as~\cite{Dedes:2015twa}
\begin{equation}\label{eq:LX}
L_\text{X}(i,j) = \sum_{n=0}^{\infty} L_\text{X}(n;i,j),
\end{equation}
where $L_\text{X}(n;i,j)$ represents the $n$-th term in the MI order of the blocks $L_\text{X}(i,j)$.

\textbf{Step 2}. During our calculation, we also encounter the case in which a Feynman diagram contains two lines involving the same particle. In such a case, the product of two blocks with MI orders specified by $m$ and $n$, $L_\text{X}(m;i,j) L_\text{X}(n;i,j)$, should be firstly combined into a single term with fixed MI order, such as  $L_{\text{UU}}(2n;3,i;3,j)=\sum_{n_1=1}^{n} L_\text{U}(2n_1-1;3,i) L_\text{U}(2n-2n_1+1;3,j)$, where $L_\text{XX}(n;i,j;i',j')$ denotes the $n$-th term in the MI order of the block $L_\text{XX}(i,j;i',j')$, with
\begin{equation}\label{eq:LXX}
L_\text{XX}(i,j;i',j') = \sum_{n=0}^{\infty} L_\text{XX}(n;i,j;i',j').
\end{equation}
All the non-zero block terms $L_\text{X}(n;i,j)$ and $L_\text{XX}(n;i,j;i',j')$ for squarks and charginos, with the flavour structures specified in Sec.~\ref{sec:FlavorStructure}, can be easily derived and are listed in the appendix.

As the neutralino mass matrix $M_{\chi_0}$, given by Eq.~\eqref{eq:Mchi0}, has many non-zero elements, one can use the following recursive formulas to represent the corresponding blocks
\begin{align}\label{eq:neublocktrans}
L_\text{N}(n;i_0,i_n)=&\sum_{i_1,i_2,\cdots,i_{n-1}} l^r_\text{N} ( i_0,i_1,\cdots,i_n )
\left(\delta^N_{{i_0}{i_1}} \sqrt{{M_N}_{i_0} {M_N}_{i_1}}\right) \nonumber\\
& \times\left(\delta^N_{{i_1}{i_2}} \sqrt{{M_N}_{i_1} {M_N}_{i_2}}\right) \cdots
\left(\delta^N_{{i_{n-1}}{i_n}} \sqrt{{M_N}_{i_{n-1}} {M_N}_{n}}\right),
\end{align}
where $l^r_\text{N}(i_0,i_1,\cdots,i_n)=\frac{1}{(q^2-{M_N}_{i_0})(q^2-{M_N}_{i_1}) \cdots (q^2-{M_N}_{i_n})}$, and can be re-expressed in terms of $l^r_\text{N}(i_x)$~$(x=0,1,\cdots,n)$ using the ``divided difference'' method~\cite{Rosiek:2015jua}.

\textbf{Step 3}. One now needs to perform the loop-momentum integration over the products of blocks $L_\text{X}(n;i,j)$, $L_\text{XX}(n;i,j;i',j')$, and $l^r_\text{N}(i_x)$ introduced in the last step. As only one-loop amplitudes are involved throughout this paper, this can be done using iteratively the operation ${\partial}_{M_{\text{SQ}}} {\text{Loop}}_\text{X}$, where $M_{\text{SQ}}$ denotes symbolically the squared mass, equaling to $M^2$ for scalars and to $M^{\dag} M$ or $M M^{\dag}$ for fermions, and ${\text{Loop}}_\text{X}$ is the $n$-point one-loop integrals in the Passarino-Veltman~(PV) basis~\cite{Passarino:1978jh}, such as $D_{2n}$, $C_{2n}$ and $B_0$ introduced in Ref.~\cite{Buras:2002vd}.

Following these three steps, one can then transform an amplitude written initially in the ME basis into an expansion in terms of the MI parameters, up to any user-defined MI order~\cite{Dedes:2015twa,Rosiek:2015jua}.

\subsection{MI-order estimation}\label{subsec:MIorder}

Although the FET procedure can provide with us the result expanded to any MI order, an optimal cutting-off should be applied due to the time costing of the programme running. Here we illustrate an efficient MI-order estimation method to get the appropriate order which was usually set by hand (order 2 or 4) in most of recent works~\cite{Dedes:2014asa,Kumar:2016vhm,Eberl:2016aox,Crivellin:2018mqz}.

Taking the block term $L_{\text{UU}}(2n;3,2,3,2)$ in case I,
\begin{align}
L_{\text{UU}}\left(2n;3,2;3,2\right)= \frac{n {M_S}_1 {M_S}_2 \delta^2_{23} }{(q^2-{M_S}_1)^2(q^2-{M_S}_2)^{n+1}}\Delta^{n-1}_{1} ,
\end{align}
where $\Delta_{1} \equiv \frac{{M_S}_1 {M_S}_2 \delta^2_{23}}{q^2-{M_S}_1}+\frac{{M^2_S}_2 \delta^2_{36}}{q^2-{M_S}_2}$, as an example, and using the inequality~\cite{Dedes:2015twa}
\begin{align}\label{eq:pv}
\left| \text{PV}_0^{(n+1)}(m^2_1,m^2_2, \cdots ,m^2_n,m^2_{n+1}) \right| \leqslant \frac{1}{m^2_{n+1}}
\left| \text{PV}_0^{(n)}(m^2_1,m^2_2, \cdots ,m^2_n) \right| ,
\end{align}
satisfied by the PV-integrals with vanishing external momenta that are defined by
\begin{align}
\text{PV}_0^{(n)}(m^2_1,m^2_2,\cdots,m^2_n)=-i(4\pi)^2 \int \frac{d^4 q}{(2\pi)^4} \frac{1}{\Pi^n_{j=1} (q^2-m_j^2)} ,
\end{align}
with the assumption that $n\geqslant3$ to avoid divergent integrals, we can obtain
\begin{align}
\left| \int d^4 q\, L_{\text{UU}}(2n;3,2;3,2)\,L_{\text{oth}} \right|
\leqslant n \left( \delta^2_{23}+\delta^2_{36} \right)^{n-1}\, \left| \int d^4 q\, L_{\text{UU}}(2;3,2;3,2)\,L_{\text{oth}} \right| ,
\end{align}
where $L_{\text{oth}}$ represent the blocks related to the other types of particles, such as $L_{\text{CC}}$. Then, for a given small constant $0<c_0<1$, only when
\begin{align}\label{LQorder}
(n+1) \left(\delta^2_{23} +\delta^2_{36} \right)^{n}<c_0,
\end{align}
can the terms starting from the $(2n+1)$-th MI order in the series expansion of the block $L_{\text{UU}}(3,2;3,2)$ be safely neglected. Thus, the cutting-off MI order for $L_{\text{UU}}(3,2;3,2)$ should be $2n$ at least. The same method can be applied for other blocks, and the final cutting-off MI orders for squarks and charginos can be determined accordingly.

For the neutralino blocks, Eq.~\eqref{eq:pv} still works for estimating the required MI order. In this case, we obtain
\begin{align}
& \left| \int d^4 q\, L_\text{N}(n;i_0,j_0)\, L_{\text{oth}} \right|
\leqslant  \sum_{i_1,i_2,\cdots,i_{n-1}}
\left| \int d^4 q\, l^r_\text{N}(i_0)\, L_{\text{oth}} \right|\,\left| \delta^N_{{i_0}{i_1}}\, \delta^N_{{i_1}{i_2}} \cdots \delta^N_{{i_{n-1}}{j_0}} \right|\, \sqrt{\frac{{M_N}_{i_0}}{{M_N}_{j_0}}} ,\\[2mm]
& \left| \int d^4 q\, L_\text{N}(n;i_0,i_n)\, {(M_{\chi_0})}_{i_n j_0}\, L_{\text{oth}} \right|
\leqslant  \sum_{i_1,i_2,\cdots,i_{n}}
\left| \int d^4 q\, l^r_N(i_0)\, L_{\text{oth}} \right| \nonumber \\
& \hspace{7.5cm}
\times \left| \delta^N_{{i_0}{i_1}}\, \delta^N_{{i_1}{i_2}} \cdots \delta^N_{{i_{n-1}}{i_{n}}}\, {(M_{\chi_0})}_{i_n j_0} \right|\,
\sqrt{\frac{{M_N}_{i_0}}{{M_N}_{i_n}}}.
\end{align}
So, when $\left| \delta^N_{{i_0}{i_1}}\, \delta^N_{{i_1}{i_2}} \cdots \delta^N_{{i_{n-1}}{j_0}} \right| < c_0$ and $\left| \delta^N_{{i_0}{i_1}}\, \delta^N_{{i_1}{i_2}} \cdots \delta^N_{{i_{n-1}}{i_{n}}}\, {(M_{\chi_0})}_{i_n j_0}\right| \frac{1}{\sqrt{{M_N}_{i_n}}} < c_0$ for fixed indices $i_0$ and $j_0$, the summation over the MI index can be terminated to the $n$-th order.

\section{\boldmath{$B_{s(d)}-\bar{B}_{s(d)}$} mixing and \boldmath{$B_s\to\mu^+\mu^-$} decay}\label{sec:3}

In this section, we shall apply the FET procedure with general/finite MI order to $B_{s(d)}-\bar{B}_{s(d)}$ mixing and $B_s \to \mu^+ \mu^-$ decay, within the $\mathbb{Z}_3$-invariant NMSSM with NMFV.

\subsection{\boldmath{$B_{s(d)}-\bar{B}_{s(d)}$} mixing}

The strength of $B_{s(d)}-\bar{B}_{s(d)}$ mixing is described by the mass difference $\Delta M_{s(d)}$, defined by~\cite{Buras:1998raa}
\begin{align}\label{eq:m12}
\Delta M_{q} = 2 | M_{12}^{q} | = 2 |\langle B_q| \mathcal H_{\rm eff}^{\Delta B=2} |\bar{B}_q\rangle|, \qquad q=s,\,d,
\end{align}
where $M_{12}^{q}$ denotes the off-diagonal element in the neutral $B_q$-meson mass matrix, and the effective weak Hamiltonian can be written in a general form as~\cite{Buras:2001ra}
\begin{align}\label{eq:Hamiltonian:mixing}
\mathcal H_{\rm eff}^{\Delta B=2} = \sum_{i} C_i Q_i + \text{h.c.}.
\end{align}
Within the $\mathbb{Z}_3$-invariant NMSSM with NMFV, the following eight operators, as defined in Ref.~\cite{Buras:2002vd}, are all found to be relevant:
\begin{align}
Q_1^{\rm VLL}&=(\bar b^\alpha\gamma_\mu P_L q^\alpha)(\bar b ^\beta \gamma^\mu P_L q^\beta) ,
& \quad
Q_1^{\rm LR}&=(\bar b^\alpha\gamma_\mu P_L q^\alpha)(\bar b^\beta\gamma^\mu P_R q^\beta) ,
\nonumber\\[2mm]
Q_1^{\rm VRR}&=(\bar b^\alpha\gamma_\mu P_R q^\alpha)(\bar b ^\beta \gamma^\mu P_R q^\beta) ,
&\quad
Q_2^{\rm LR}&=(\bar b^\alpha P_L q^\alpha)(\bar b^\beta P_R q^\beta) ,
\nonumber\\[2mm]
Q_1^{\rm SLL}&=(\bar b^\alpha P_L q^\alpha)(\bar b^\beta P_L q^\beta) ,
&\quad
Q_2^{\rm SLL}&=(\bar b^\alpha \sigma_{\mu\nu} P_L q^\alpha)(\bar b ^\beta \sigma^{\mu\nu} P_L q^\beta) ,
\nonumber\\[2mm]
Q_1^{\rm SRR}&=(\bar b^\alpha P_R q^\alpha)(\bar b^\beta P_R q^\beta) ,
&
\quad Q_2^{\rm SRR}&=(\bar b^\alpha \sigma_{\mu\nu} P_R q^\alpha)(\bar b ^\beta \sigma^{\mu\nu} P_R q^\beta) ,
\end{align}
where $\alpha$ and $\beta$ are the colour indices, $\sigma_{\mu\nu}=\frac{1}{2}[\gamma_\mu,\gamma_\nu]$, and $P_{L,R}=(1\mp\gamma_5)/2$.

To the lowest order in the EW theory, the corresponding Wilson coefficients $C_i$, at the matching scale, of the operators $Q_i$ are obtained by evaluating the various one-loop box diagrams mediated by heavy particles appearing in the SM and beyond\footnote{Here we do not consider the double-penguin diagrams, which involve the exchange of CP-even and CP-odd scalars, and can give significant contributions only for large values of $\tan\beta$~\cite{Buras:2002vd,Hamzaoui:1998nu,Dedes:2003kp}. This is justified by our choices of the two sets of SUSY parameters collected in Table~\ref{tab:susyinputs}, with $\tan\beta$ being fixed at $3$ and $10$, respectively.}. Within the SM, only $C_1^{\rm VLL}$ gets a non-negligible contribution from the one-loop box diagrams with up-type quarks and $W$ bosons circulating in the loops~\cite{Buras:1998raa}, and the next-to-leading-order perturbative QCD corrections to $C_1^{\rm VLL}$ are also known~\cite{Buras:1990fn}. In the context of $\mathbb{Z}_3$-invariant NMSSM with NMFV, on the other hand, all the eight Wilson coefficients $C_i$ can get non-zero contributions from the additional one-loop box diagrams mediated by: 1) charged Higgs, up-quarks; 2) chargino, up-squarks; 3) gluinos, down-squarks; 4) neutralinos, down-squarks; 5) mixed gluino, neutralino, down-squarks~\cite{Altmannshofer:2007cs,Bertolini:1990if,Buras:2002vd}. With the aid of the packages {\tt FeynArts}~\cite{Hahn:2000kx} and {\tt FeynCalc}~\cite{Shtabovenko:2016sxi}, all these Feynman diagrams can be calculated and the resulting Wilson coefficients are expressed in terms of the rotation matrices $Z_U$, $Z_D$, $Z_{\chi}^+$, $Z_{\chi}^-$, and $Z_{\chi_0}$, as well as the blocks $L_\text{X}(i,j)$ and $L_\text{XX}(i,j;i',j')$. Our results for the Wilson coefficients agree with the ones given in Refs.~\cite{Kumar:2016vhm,AranaCatania:2011ak,Altmannshofer:2007cs}. Then, following the procedure detailed in Sec.~\ref{subsec:HMIFETpro}, we can transform these Wilson coefficients given in the ME basis into the FET results. Here we have made full use of the hierarchies among the CKM parameters to simplify the final results. For example, when calculating $\Delta M_{s}$ in case I, we encounter a term
\begin{align}
\sum_{i,j} K^{*}_{i3}L_{\text{UU}}(i,j)K_{j2} = \sum_{i} K^{*}_{i3}L_{\text{UU}}(i,i)K_{i2}+\sum_{i\neq j} K^{*}_{i3}L_{\text{UU}}(i,j)K_{j2}.
\end{align}
As $L_{\text{UU}}(i,j)$, with $i\neq j$, does not vanish only when $(i,j)=(2,3)$ or $(3,2)$, and because of $|K^{*}_{23} K_{32}|\ll|K^{*}_{33} K_{22}|$, we can neglect safely the term with $(i,j)=(2,3)$, to get
\begin{align}
\sum_{i,j} K^{*}_{i3}L_{\text{UU}}(i,j)K_{j2} \approx \sum_{i} K^{*}_{i3}L_{\text{UU}}(i,i)K_{i2}+ K^{*}_{33}L_{\text{UU}}(3,2)K_{22}.
\end{align}

Once the initial conditions for the Wilson coefficients are obtained, we need to include the renormalization group~(RG) running effects from the matching scale down to the low-energy scale, at which the hadronic matrix elements are evaluated by the lattice groups~\cite{Bazavov:2016nty,Aoki:2019cca}. All the relevant ingredients for this RG running can be found in Ref.~\cite{Buras:2001ra}. In this way, we can obtain the final result of the off-diagonal element $M_{12}^{q}$ and hence the mass difference $\Delta M_{s(d)}$. For convenience of later discussions, the total contributions to $M_{12}^{q}$ are split into the following different parts:
\begin{equation}\label{eq:m12decomposition}
M_{12}^{q} \equiv M_{12}^{(q)\,\text{SM}}+M_{12}^{(q)\,\text{CH}}+M_{12}^{(q)\,\text{C}}+M_{12}^{(q)\,\text{NG}},
\end{equation}
where $M_{12}^{(q)\,\text{SM}}$, $M_{12}^{(q)\,\text{CH}}$, $M_{12}^{(q)\,\text{C}}$, and $M_{12}^{(q)\,\text{NG}}$ represent contributions from the SM, the charged Higgs, the charginos, as well as the neutralinos and gluinos, respectively.

\subsection{$\boldsymbol{B_s\to \mu^+ \mu^-}$ decay}

The rare decay $B_s\to \mu^+ \mu^-$ proceeds dominantly via the $Z$-penguin and $W$-box diagrams within the SM and its branching ratio is highly suppressed~\cite{Buras:1998raa}. In the context of $\mathbb{Z}_3$-invariant NMSSM with NMFV, there are in general three types of one-loop diagrams that contribute to this decay, including the various box, the $Z$-penguin, and the neutral-Higgs-penguin diagrams~\cite{Buras:2002vd,Bobeth:2001jm,Bobeth:2001sq,Chankowski:2000ng,Huang:2000sm,Isidori:2002qe,Dedes:2008iw,Dreiner:2012dh}. The relevant effective weak Hamiltonian reads~\cite{Dedes:2008iw,Dreiner:2012dh}\footnote{Here we need not consider the tensor operators $\mathcal O_{TX}=(\bar b \sigma^{\mu \nu} P_X s)(\bar{\mu} \sigma_{\mu \nu} \mu )$. While also receiving contributions from these three types of one-loop diagrams in the $\mathbb{Z}_3$-invariant NMSSM with NMFV, they do not contribute to this process due to the vanishing matrix elements, $\langle 0|\bar b \sigma^{\mu \nu} P_X s|B_s\rangle=0$.}
\begin{align}\label{eq:Hamiltonian}
\mathcal H_{\rm eff}
=\frac{1}{16 \pi^2} \sum_{X,Y=L,R}  \big(C_{VXY}\mathcal O_{VXY}+C_{SXY}\mathcal O_{SXY}\big)+ \text{h.c.},
\end{align}
where the vector~($\mathcal O_{VXY}$) and scalar~($\mathcal O_{SXY}$) operators are defined, respectively, by
\begin{align}\label{eq:operator}
\mathcal O_{VXY}=(\bar b \gamma_\mu P_X s)(\bar{\mu} \gamma^\mu P_Y \mu ) , \qquad
\mathcal O_{SXY}=(\bar b P_X s)(\bar{\mu} P_Y \mu ).
\end{align}
The branching ratio of $B_s \to \mu^+ \mu^-$ decay is then calculated to be~\cite{Dedes:2008iw,Dreiner:2012dh}
\begin{align}\label{BRbqlltheo}
\mathcal B (B_s \to \mu^+\,\mu^-)=
\frac{\tau_{B_s}}{16 \pi} \frac{|\mathcal M|^2}{M_{B_s}} \sqrt{1-{\left( \frac{2 m_{\mu}}{M_{B_s}} \right)}^2} ,
\end{align}
where $\tau_{B_s}$ is the $B_s$-meson lifetime, and the squared matrix element is given by~\cite{Dedes:2008iw,Dreiner:2012dh,Bobeth:2002ch,Isidori:2002qe}
\begin{align}
(4 \pi)^4 |\mathcal M|^2=&2|F_S|^2 \left(M_{B_s}^2-4 m_{\mu}^2 \right)+2|F_P|^2 M_{B_s}^2
+8|F_A|^2 M_{B_s}^2 m_{\mu}^2
+8\text{Re}\left(F_P F_A^*\right) M_{B_s}^2 m_{\mu} ,
\end{align}
with the scalar, pseudo-scalar, and axial-vector form factors defined, respectively, by~\cite{Dedes:2008iw}
\begin{align}
F_S&=\frac{i}{4}\frac{M_{B_s}^2 f_{B_s}}{\overline{m}_b+\overline{m}_s}(C_{SLL}+C_{SLR}-C_{SRR}-C_{SRL}) ,\\[2mm]
F_P&=\frac{i}{4}\frac{M_{B_s}^2 f_{B_s}}{\overline{m}_b+\overline{m}_s}(-C_{SLL}+C_{SLR}-C_{SRR}+C_{SRL}) ,\\[2mm]
F_A&=-\frac{i}{4}f_{B_s}(-C_{VLL}+C_{VLR}-C_{VRR}+C_{VRL}),
\end{align}
where $f_{B_s}$ is the $B_s$-meson decay constant, and $\overline{m}_{b(s)}$ denotes the $b(s)$-quark running mass.

Our main task is then to calculate the Wilson coefficients $C_{VXY}$ and $C_{SXY}$. Within the SM, only $C_{VLL}$ gets a non-negligible contribution\footnote{When the small external momenta and masses are considered, the SM $W$-box and $Z$-penguin diagrams also generate non-zero contributions to $C_{SXY}$, besides the ones from the Higgs-penguin diagrams~\cite{Li:2014fea,Arnan:2017lxi}.}, and we shall use the fitting formula in Eq.~(4) of Ref.~\cite{Bobeth:2013uxa} to get the numerical result for it. The additional contributions to $C_{VXY}$ and $C_{SXY}$ from the $\mathbb{Z}_3$-invariant NMSSM with NMFV are calculated by ourselves, with the aid of {\tt FeynArts} and {\tt FeynCalc} packages. Then, the FET procedure with general/finite MI order is applied for the NMSSM contributions, in exactly the same way as for the $B_{s(d)}-\bar{B}_{s(d)}$ mixing.

It should be noted that the branching ratio given by Eq.~\eqref{BRbqlltheo} is the so-called ``theoretical'' branching ratio, which corresponds to the decay time $t=0$, while the experimentally measured branching ratio is time-integrated, which is given by~\cite{DeBruyn:2012wk,Buras:2013uqa,Fleischer:2017ltw}
\begin{align}
\overline{\mathcal B}(B_s \to \mu^{+}\mu^{-}) = \frac{1+A_{\Delta {\Gamma}_s} y_s}{1-y_s^2}\,
{\mathcal B}(B_s \to \mu^{+}\mu^{-}) ,
\end{align}
where $A_{\Delta {\Gamma}_s}$ is a time-dependent observable~\cite{DeBruyn:2012wk,Buras:2013uqa,Fleischer:2017ltw}, and $y_s$ is related to the decay-width difference $\Delta \Gamma_s$ between the two $B_s$-meson mass eigenstates, defined by
\begin{align}
y_s\equiv\frac{\Gamma^s_L-\Gamma^s_H}{\Gamma^s_L+\Gamma^s_H}=\frac{\Delta \Gamma_s}{2 \Gamma_s} ,
\end{align}
with $\Gamma^s_L$ ($\Gamma^s_H$) denoting the lighter (heavier) eigenstate decay width and $\Gamma_s=\tau_{B_s}^{-1}$  the average decay width of $B_s$ meson. In the absence of beyond-SM sources of CP violation, which is assumed throughout this paper, both $A_{\Delta {\Gamma}_s}$ and $y_s$ will take their respective SM values~\cite{DeBruyn:2012wk,Buras:2013uqa,Fleischer:2017ltw,Altmannshofer:2017wqy}, and the two branching ratios are then related to each other via a simple relation
\begin{align}
\overline{\mathcal B} (B_s \to \mu^{+}\mu^{-}) = \frac{1}{\tau_{B_s}\Gamma^s_H}\,\mathcal B(B_s \to \mu^{+}\mu^{-}),
\end{align}
which holds to a very good approximation~\cite{Bobeth:2013uxa}.

\section{Numerical results and discussions}\label{sec:4}

After getting the analytic FET results for the $B_{s(d)}-\bar{B}_{s(d)}$ mixing and $B_s \to \mu^+ \mu^-$ decay, we now proceed to analyze numerically the parameter space of $\mathbb{Z}_3$-invariant NMSSM with NMFV that is allowed under these experimental constraints.

\subsection{Choice of input parameters}

\begin{table}[t]
	\tabcolsep 0.42in
	\renewcommand\arraystretch{1.25}
	\begin{center}
		\caption{\label{tab:inputs} \small Summary of part of the input parameters used throughout this paper.}
		\vspace{0.18cm}
		\begin{tabular}{|c|c|c|c|}
			\hline\hline
			\multicolumn{4}{|l|}{QCD and EW parameters~\cite{Tanabashi:2018oca}}\\
			\hline
			$G_F[10^{-5}~\text{GeV}^{-2}]$ & $\alpha_s(M_Z)$  &  $M_W[\text{GeV}]$ & $\sin^2\theta_W$\\
			\hline
			$1.1663787$ & $0.1181(11)$ &  $80.379$ & $0.2312$\\
			\hline
			\multicolumn{4}{|l|}{Quark masses [GeV]~\cite{Tanabashi:2018oca}}\\
			\hline
			$\overline{m}_b(\overline{m}_b)$ & $\overline{m}_c(\overline{m}_c)$ & \multicolumn{2}{c|}{$m_t$}\\
			\hline
			$4.18^{+0.04}_{-0.03}$ & $1.275^{+0.025}_{-0.035}$ & \multicolumn{2}{c|}{$173.0(4)$}  \\
			\hline
			\multicolumn{4}{|l|}{$B$-meson parameters~\cite{Amhis:2016xyh}}\\
			\hline
			$M_{B_d}[\text{GeV}]$  & $M_{B_s}[\text{GeV}]$ & \multicolumn{2}{c|}{$1/\Gamma^s_H[\text{ps}]$}\\
			\hline
			$5.280$ & $5.367$ & \multicolumn{2}{c|}{$1.609(10)$} \\
			\hline
			\multicolumn{4}{|l|}{CKM parameters~\cite{Koppenburg:2017mad}}\\
			\hline
			$\lambda_{\text{CKM}}$ & $A$ & $\overline{\rho}$ & $ \overline{\eta}$ \\
			\hline
			$0.2251(4)$ & $0.831^{+0.021}_{-0.031}$ & $0.155(8)$ & $0.340(10)$ \\
			\hline \hline
		\end{tabular}
	\end{center}
\end{table}

Firstly, we collect in Table~\ref{tab:inputs} part of the input parameters used throughout this paper. For the $B_{s(d)}$-mixing matrix elements and the decay constants $f_{B_{s(d)}}$, we take the values provided by the FNAL/MILC collaboration~\cite{Bazavov:2016nty} and the averages by the Particle Data Group~\cite{Tanabashi:2018oca}, respectively. The relevant model parameters of the $\mathbb{Z}_3$-invariant NMSSM with NMFV include
\begin{align}
M_1,\ M_2, \ M_{\tilde{g}},\ {M_S}_1,\ {M_S}_2,\ \mu_\text{eff},\ \tan \beta,\ A_{\lambda},\ A_{\kappa},\ \lambda,\ \kappa,\ \delta_{23}^{\text{LL}},\ \delta_{33}^{\text{LR}},\ \delta_{13}^{\text{RR}},\ \delta_{23}^{\text{RR}},
\end{align}
where $M_{\tilde{g}}$ is the gluino mass. In this paper, we shall consider two sets of fixed parameters that are collected in Table~\ref{tab:susyinputs}. Scenario A is characterized by a large $\lambda$ and a small $\tan\beta$, to avoid suppressing the NMSSM-specific tree-level contributions to the SM-like Higgs mass~\cite{Ellwanger:2009dp,Maniatis:2009re}, while scenario B is featured by negligible NMSSM-specific effects on the SM-like Higgs mass~\cite{Staub:2015aea}. The remaining parameters ${M_S}_2$,  $\delta_{23}^{\text{LL}}$, $\delta_{33}^{\text{LR}}$, $\delta_{13}^{\text{RR}}$, and $\delta_{23}^{\text{RR}}$ can vary freely in the ranges considered.
\begin{table}[t]
	\tabcolsep 0.12in
	\renewcommand\arraystretch{1.25}
	\begin{center}
		\caption{\label{tab:susyinputs} \small Two set of fixed parameters, all being defined at the scale $1~\rm{TeV}$, for the $\mathbb{Z}_3$-invariant NMSSM with NMFV. They are given in units of ``GeV'' except for $\tan\beta$, $\lambda$, and $\kappa$. }
		\vspace{0.18cm}
		\begin{tabular}{c c c c c c c c c c c }	
			\hline\hline
			& $M_1$ & $M_2$  & $M_{\tilde{g}}$ & $\ \sqrt{{M_S}_1}$ & $\mu_\text{eff}$ & $\tan \beta$ & $A_{\lambda}$ & $A_{\kappa}$ & $\lambda$ & $\kappa$  \\ \hline
			Scenario \textrm{A} & $500$& $1000$& $2100$ & $1600$ & $200$ & $3$ & $650$ & $-10$ & $0.67$ & $0.1$   \\ \hline
			Scenario \textrm{B} &$500$& $1000$& $2100$ & $1600$ & $200$ & $10$ & $2000$ & $-100$ & $0.3$ & $0.2$ \\
			\hline \hline
		\end{tabular}
	\end{center}
\end{table}

The set of fixed parameters in scenario~\textrm{A} is similar to that of the scenario TP3 in Ref.~\cite{Staub:2015aea}, with $\lambda$ being close to the perturbative limit but still avoiding running into a Landau pole well below the GUT scale~\cite{Ellwanger:2009dp}. The one in scenario~\textrm{B} is, however, featured by a large $A_{\lambda}$, which is closely related to the charged-Higgs mass~\cite{Ellwanger:2009dp}. In scenario~\textrm{A}, the predicted lightest and next-to-lightest neutralino masses are given, respectively, by $m_{\tilde{\chi}_1^0}\sim74$~GeV and $m_{\tilde{\chi}_2^0}\sim211$~GeV, which comply with the limits set by the most recent search for electroweak production of charginos and neutralinos, via the most promising channel $p p\to \tilde{\chi}_1^{\pm}\,\tilde{\chi}_2^0$ with a $100\%$ branching fraction of $\tilde{\chi}_2^0\to H_{\text SM}\,\tilde{\chi}_1^0$, by the CMS collaboration~\cite{Sirunyan:2018ubx}. In scenario~\textrm{B}, on the other hand, $m_{\tilde{\chi}_1^0}\sim 179$~GeV and $m_{\tilde{\chi}_2^0}\approx m_{\tilde{\chi}_1^{\pm}}\sim 207$~GeV, being compatible with the mass bounds from the same channel but with a $100\%$ branching fraction of $\tilde{\chi}_2^0\to Z\,\tilde{\chi}_1^0$~\cite{Sirunyan:2018ubx}. The mass splitting $\Delta m(\tilde{\chi}_2^0,\tilde{\chi}_1^0)=m_{\tilde{\chi}_2^0}-m_{\tilde{\chi}_1^0}$ in both of these two scenarios is also compatible with the bound set by the CMS collaboration~\cite{Sirunyan:2018iwl,Sirunyan:2018ubx}. In addition, the choice $\mu_\text{eff} = 200$~GeV not only complies with the lower bounds on chargino and neutralino masses, but also ensures that no tree-level fine-tuning is necessary to achieve the EW symmetry breaking~\cite{Ellwanger:2009dp}.

The LHC direct searches have also led to stringent limits on the masses of stops, sbottoms, and gluinos~\cite{Sirunyan:2017leh,Aaboud:2017phn,Aad:2015gna,Sirunyan:2017xse,Sirunyan:2017pjw,Sirunyan:2017kiw,Aaboud:2017ayj,Aaboud:2017aeu}. The recent ATLAS result~\cite{Aaboud:2017aeu} has shown that, for pair produced stops decaying into top quarks, stop masses up to $940$~GeV are already excluded in the phenomenological MSSM with a wino-like next-to-lightest supersymmetric particle, while the excluding limit can be up to $860$~GeV in scenarios with a Higgsino-like lightest supersymmetric particle. Obviously, being based on simplified models, these bounds are obtained without considering the most general squark flavour structures, and can be relaxed when taking into account these mixings~\cite{Brooijmans:2018xbu}. Furthermore, there exist possible NMSSM-specific effects with a light singlino in the searches for squarks, which could modify these limits~\cite{Ellwanger:2014hia,Kim:2015dpa,Allanach:2015mwa,Potter:2015wsa,Beuria:2016mur,Titterton:2018pba}. Here we simply assume that the mass of the lightest squark is above $940$~GeV, and fix the gluino mass at $2100$~GeV~\cite{Ellwanger:2014dfa}.

Scenario~\textrm{A} will give a light singlet scalar with mass around $90$~GeV and a $125$~GeV SM-like Higgs, while the charged-Higgs mass is around $650$~GeV, which may be beneficial for describing the branching ratio of $B\rightarrow X_s \gamma$ decay~\cite{Hu:2016gpe,Misiak:2017bgg}. The SM-like Higgs predicted in scenario~\textrm{B} is, on the other hand, the lightest among the neutral scalars, and the charged Higgs with mass around $2$~TeV can make its effect on the $B\rightarrow X_s \gamma$ decay marginal. Taking together with the values of $\tan\beta$, we can say that both of these two scenarios make the Higgs-penguin effects negligible for both $B_{s(d)}-\bar{B}_{s(d)}$ mixing~\cite{Kumar:2016vhm} and $B_s \to \mu^+ \mu^-$ decay~\cite{Dedes:2008iw}.

\begin{figure}[ht]
	\centering
	\includegraphics[width=0.98\textwidth]{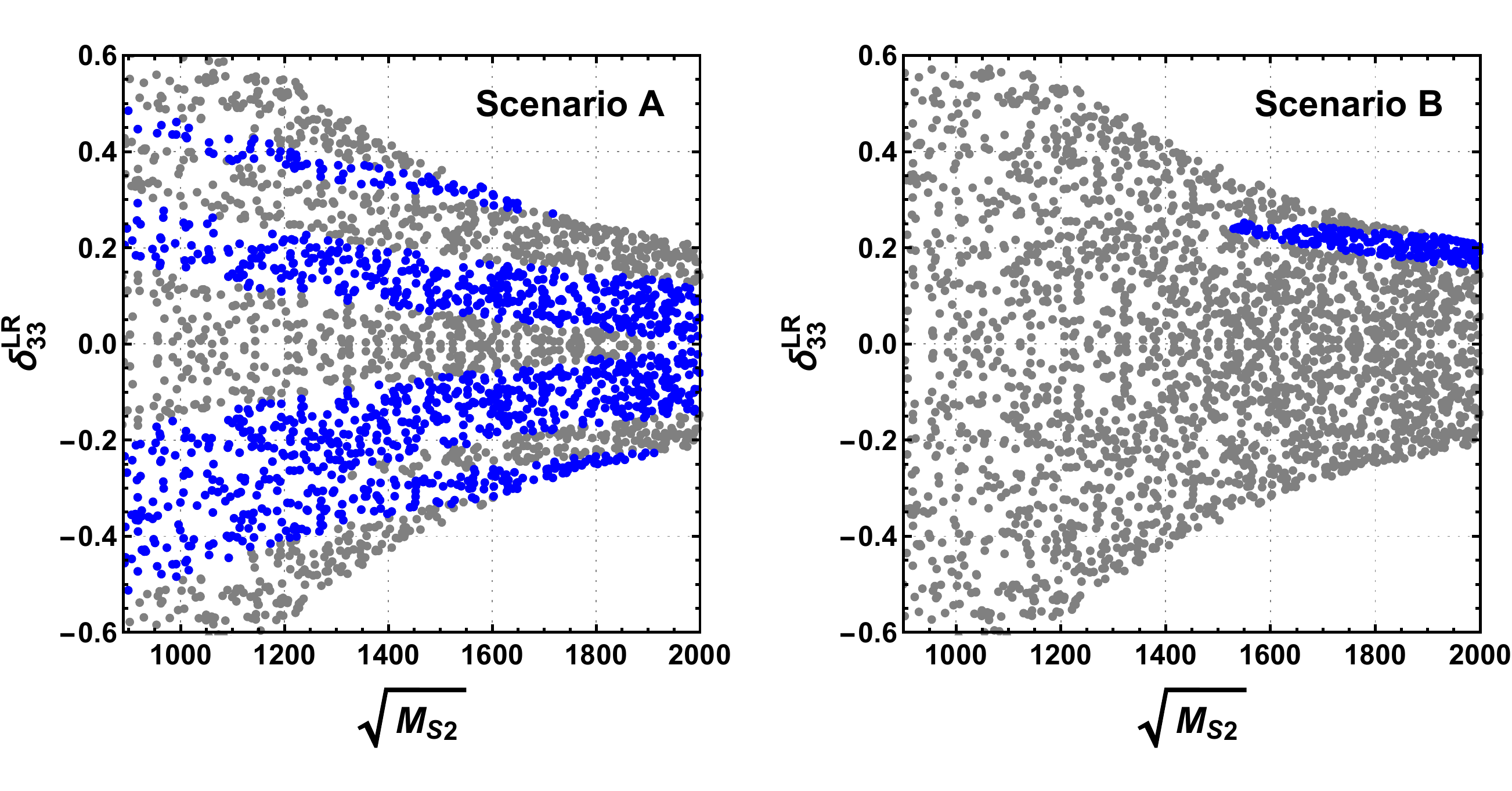}
	\caption{\label{higgs125} \small Allowed regions (blue dots) for the parameters ${M_S}_2$ and $\delta_{33}^{\text{LR}}$ required to match the SM-like Higgs mass in scenarios \textrm{A} (left) and \textrm{B} (right), respectively. The gray dots indicate that the SM-like Higgs mass is not in the range $122~{\rm GeV}\leqslant m_{H_{SM}}\leqslant 128~{\rm GeV}$.}
\end{figure}

The parameters $\delta_{33}^{\text{LR}}$ and ${M_S}_2$ are chosen to get the SM-like Higgs mass in the range $122~{\rm GeV}\leqslant m_{H_{SM}}\leqslant 128~{\rm GeV}$~\cite{Cao:2016nix}. The allowed regions for $\delta_{33}^{\text{LR}}$ and ${M_S}_2$, obtained with the aid of the package {\tt NMSSMCALC}~\cite{Baglio:2013iia,Ender:2011qh,Graf:2012hh,Muhlleitner:2014vsa,King:2015oxa,Djouadi:1997yw,Butterworth:2010ym}, are shown in Figure~\ref{higgs125}. It can be seen that $940^2~\text{GeV}^2\leqslant {M_S}_2\leqslant 2000^2~\text{GeV}^2$ and $-0.5\leqslant \delta_{33}^{\text{LR}}\leqslant 0.5$ in scenario~\textrm{A}, while $1500^2~\text{GeV}^2\leqslant {M_S}_2\leqslant 2000^2~\text{GeV}^2$ and $\delta_{33}^{\text{LR}}$ is only allowed to be around $0.2$ in scenario~\textrm{B}. These bounds will be taken into account in the following numerical analyses.

\subsection{FET result with optimal MI order}

\subsubsection{Cutting-off MI-order estimation}\label{sssec:OE}

\begin{figure}[ht]
	\centering
	\includegraphics[width=0.96\textwidth]{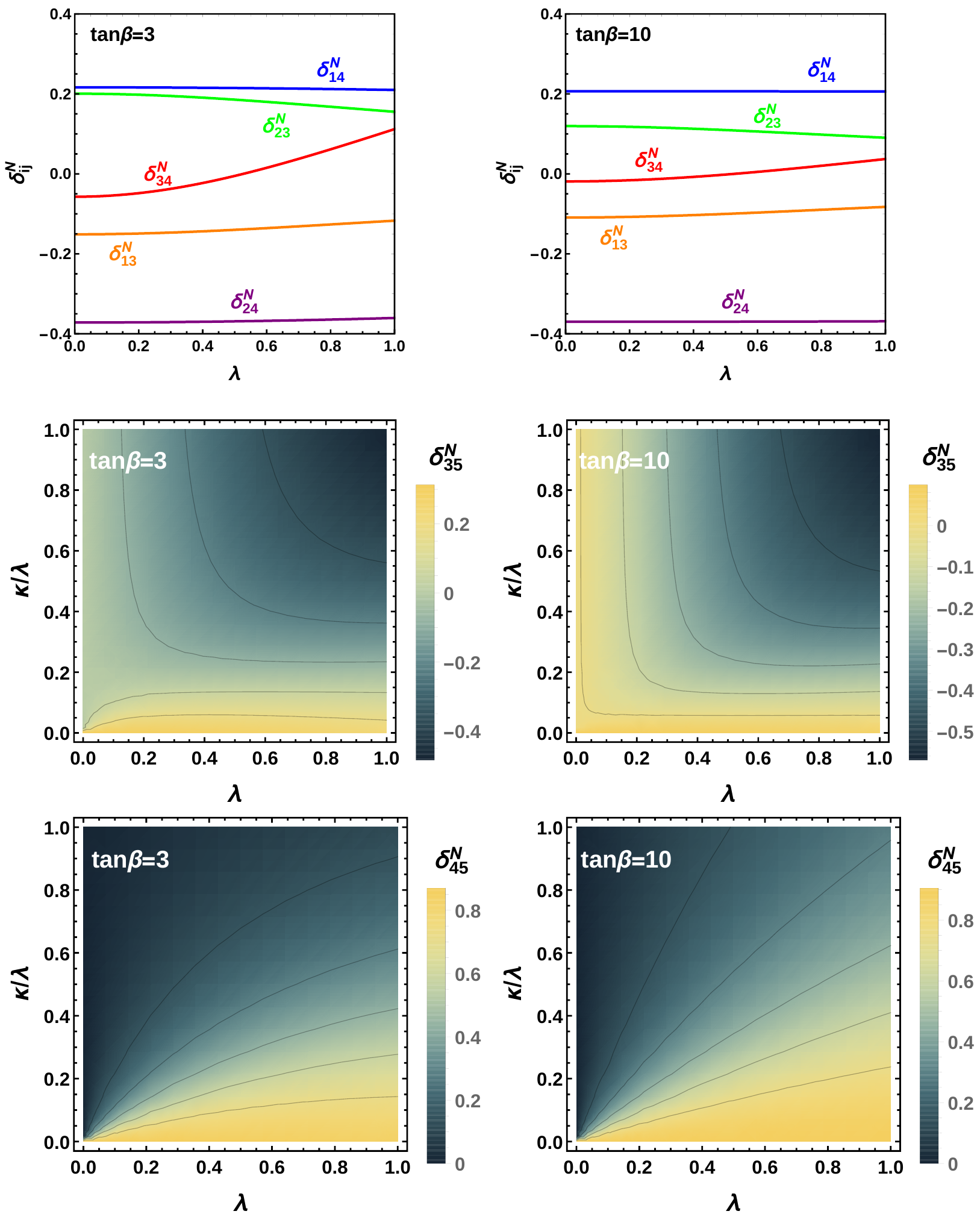}
	\caption{\label{DelN} \small Dependence of the MI parameters $\delta^N_{ij}$ for neutralinos on the $\mathbb{Z}_3$-invariant NMSSM inputs $\lambda$ and $\kappa$, with $\tan\beta$ fixed at $3$ (left panel) and $10$ (right panel), respectively.}
\end{figure}

We firstly estimate the optimal cutting-off MI order for neutralinos. Here the MI parameters $\delta^N_{ij}$ depend on the six $\mathbb{Z}_3$-invariant NMSSM parameters $M_1$, $M_2$, $\mu_\text{eff}$, $\tan\beta$, $\lambda$, and $\kappa$. Keeping for the moment $\lambda$ and $\kappa$ as free variables, but with $\tan\beta=3$ and $10$ corresponding respectively to the two scenarios defined in Table~\ref{tab:susyinputs}, we find that $\delta^N_{15}=\delta^N_{25}=0$ and $\delta^N_{12}=-0.007$. The dependence of the other MI parameters $\delta^N_{ij}$ on $\lambda$ and $\kappa$ are displayed in Figure~\ref{DelN}. From the magnitudes of these $\delta^N_{ij}$ shown and based on the criterion specified in Sec.~\ref{subsec:MIorder}, one can see that the effects of $\delta^N_{12}$, $\delta^N_{13}$, and $\delta^N_{34}$ are negligible with the MI order higher than $1$, and that of $\delta^N_{14}$ and $\delta^N_{23}$ can be neglected starting from the third MI order (even from the second MI order for $\delta^N_{23}$ in scenario~\textrm{B}); while the MI orders higher than $3$ should be kept for $\delta^N_{24}$, $\delta^N_{35}$, and $\delta^N_{45}$. When the parameters $\lambda$ and $\kappa$ are fixed at the values in scenarios~\textrm{A} and \textrm{B}, it is further found that the values of ${{(M_{\chi_0})}_{i_n\,3}}/{\sqrt{{M_N}_{i_n}}}$, for $i_n=1,\,2,\,3$, are much smaller than for $i_n=4,\,5$, and all the terms involving $L_\text{N}(n;3,i_n) {(M_{\chi_0})}_{i_n 3}$ can be, therefore, discarded safely for $i_n=1,\,2,\,3$.

As the MI parameters for charginos are all less than $0.6$ in scenarios~\textrm{A} and \textrm{B}, the optimal cutting-off MI order for chargino and squark is determined to be $8$ using Eq.~\eqref{LQorder}, when the squark MI parameters are less than $0.6$ and the given small parameter $c_0$ is set to be $0.1$.

\subsubsection{MI-order comparison for $\boldsymbol{B_{s(d)}-\bar{B}_{s(d)}}$ mixing}

\begin{figure}[htbp]
	\centering
	\includegraphics[width=0.98\textwidth]{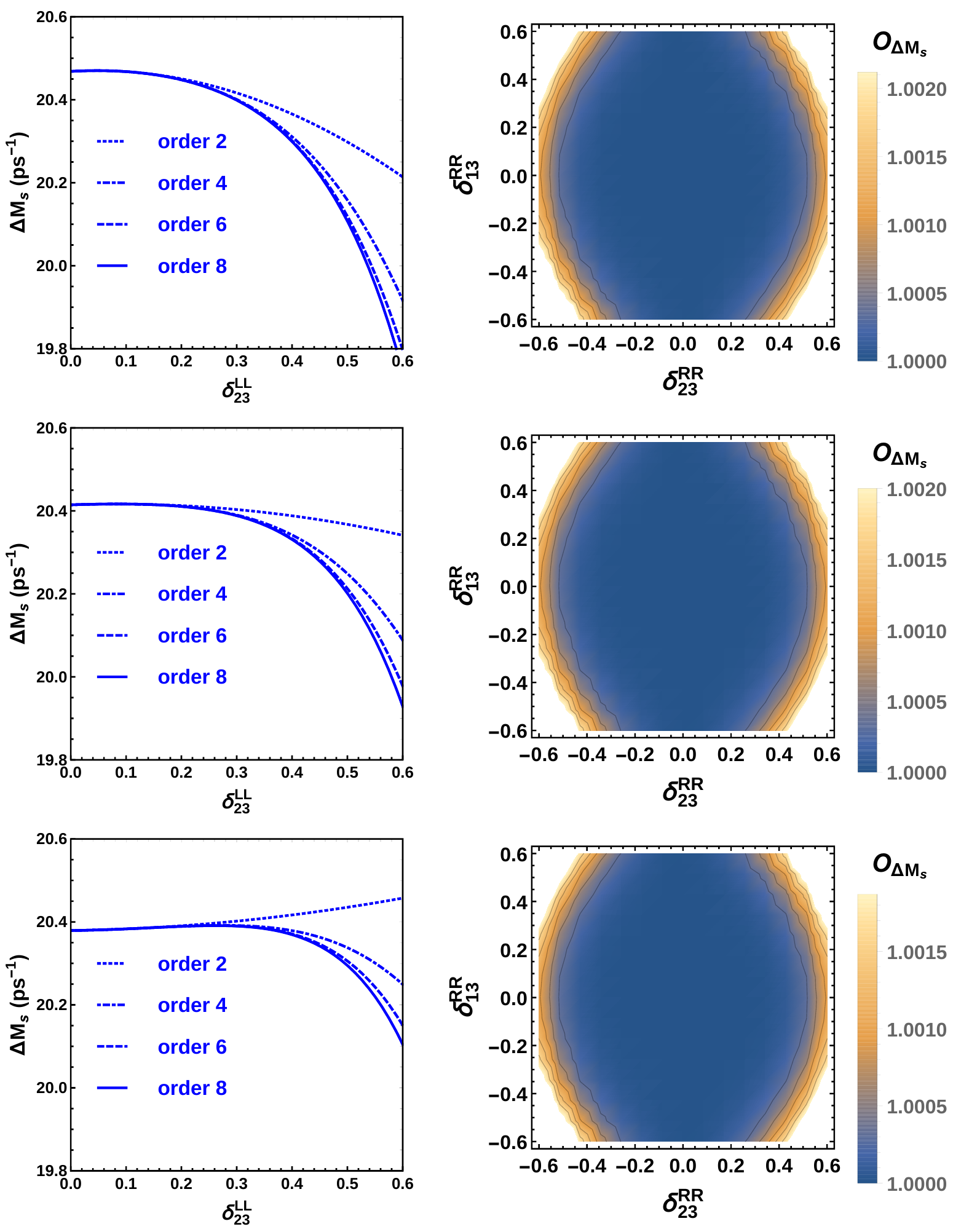}
	\caption{\label{fig:TP3REmsq1500vde} \small Variations of $\Delta M_s$ with respect to $\delta_{23}^{\text{LL}}$ in case~\textrm{IA} (left panel) and $O_{\Delta M_s}$ with respect to $\delta_{23}^{\text{RR}}$ and $\delta_{13}^{\text{RR}}$ in case~\textrm{IIA} (right panel). Here $\sqrt{{M_S}_2}$ is fixed at $1100$ (upper), $1300$ (middle), and $1500~\text{GeV}$ (lower) in both cases. In case~\textrm{IA}, $\delta_{33}^{\text{LR}}$ is set to be $-0.15$ by considering the cutting-off MI orders of $2$ (dotted blue), $4$ (dot-dashed blue), $6$ (dashed blue), and $8$ (solid blue), respectively.}
\end{figure}

After obtaining the optimal cutting-off MI orders, we now check the convergence of the FET results with different MI orders. Let us firstly discuss the $B_{s(d)}-\bar{B}_{s(d)}$ mixing. The left panel in Figure~\ref{fig:TP3REmsq1500vde} shows the variation of $\Delta M_s$ with respect to $\delta_{23}^{\text{LL}}$ by considering the cutting-off MI order for squarks from $2$ to $8$ in case~\textrm{IA}, in which the set of fixed parameters in scenario~\textrm{A} is used under the case I assumption for the squark flavour structures (and similar definitions apply to the cases \textrm{IB}, \textrm{IIA}, and \textrm{IIB}, to be mentioned below). Here the results with MI order of $2$ are similar to what are usually considered in the MIA method. One can see that $\Delta M_s$ varies with respect to $\delta_{23}^{\text{LL}}$ only slowly for MI order of $2$ but decreases obviously for higher MI orders. The results with MI orders of $6$ and $8$ are nearly identical, which justifies the validity of our MI-order estimation, and hence the cutting-off MI order $8$ or $6$ is optimal for the considered range of $\delta_{23}^\text{LL}$.

In order to show the convergence of the FET results in the case II assumption for squark flavour structures, we consider the ratio
\begin{align}
O_{\Delta M_s}= \frac{\Delta M_s \big|_{\text{MI-order}=6}}{\Delta M_s \big|_{\text{MI-order}=8}},
\end{align}
which gives the difference between $\Delta M_s$ by considering the squark MI orders of $6$ and $8$, respectively. As shown in the right panel of Figure~\ref{fig:TP3REmsq1500vde}, $O_{\Delta M_s}$ is nearly $1$ in the whole area in case~\textrm{IIA}, implying that the convergence has been verified. Similar observations are also made for $\Delta M_s$ in case~\textrm{IIB} and for $\Delta M_d$ in all the four cases.

\subsubsection{MI-order comparison for $\boldsymbol{B_s\to \mu^+ \mu^-}$ decay}

\begin{figure}[ht]
	\centering
	\includegraphics[width=0.99\textwidth]{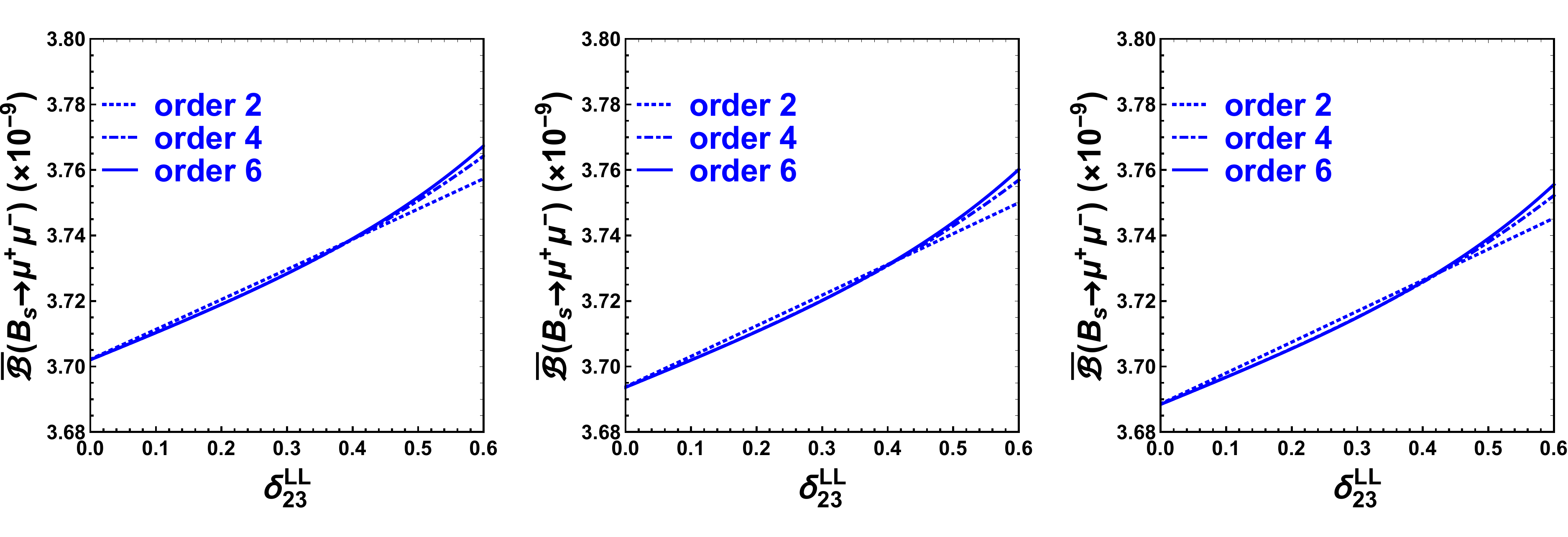}
	\caption{\label{TP3REdel23difAtBRS} \small Variation of $\overline{\mathcal B}(B_s \to \mu^+ \mu^-)$ with respect to $\delta_{23}^{\text{LL}}$ for $\delta_{33}^{\text{LR}}=-0.15$ in case~\textrm{IA}. Here $\sqrt{{M_S}_2}$ is fixed at $1100$ (left), $1300$ (middle), and $1500~\text{GeV}$ (right), respectively. The dotted, dot-dashed, and solid blue curves represent the results with squark cutting-off MI order of $2$, $4$, and $6$, respectively.}
\end{figure}

For our prediction of the time-integrated branching ratio $\overline{\mathcal B}(B_s \to \mu^+ \mu^-)$ in the $\mathbb{Z}_3$-invariant NMSSM with NMFV, the slepton mass squared matrices are set to be diagonal, with all diagonal elements being given by ${M_S}_1$. As an example of convergence checking, we show in Figure~\ref{TP3REdel23difAtBRS} the variation of $\overline{\mathcal B}(B_s \to \mu^+ \mu^-)$ with respect to $\delta_{23}^{\text{LL}}$ for $\delta_{33}^{\text{LR}}=-0.15$ in case~\textrm{IA}. Being affected by $\delta_{13}^{\text{RR}}$ and $\delta_{23}^{\text{RR}}$ quite weakly~\cite{Silvestrini:2007yf,Kumar:2016vhm}, the convergence of $\overline{\mathcal B}(B_s \to \mu^+ \mu^-)$ in case II is not shown here. It can be seen from Figure~\ref{TP3REdel23difAtBRS} that the FET result obtained with squark MI order of $2$ has no significant deviation from that with higher MI orders, which is obviously different from what is observed in the case of $B_{s(d)}-\bar{B}_{s(d)}$ mixing.

When the off-diagonal element $\delta_{23}^{\text{LL}}$ is zero, there exists no squark MI contribution and the NMSSM contributions come only from the charged Higgs and the diagonal part of $M_{\tilde{U}}^2$. Even in this case, $\overline{\mathcal B}(B_s \to \mu^+ \mu^-)$ can be obviously enhanced, putting it to be larger than the LHCb measurement, $(3.0\pm0.6^{+0.3}_{-0.2})\times10^{-9}$~\cite{Aaij:2017vad}. As a reference, our prediction within the SM is $(3.6 \pm 0.3)\times10^{-9}$, obtained by using the fitting formula in Eq.~(4) of Ref.~\cite{Bobeth:2013uxa} with the updated input parameters listed in Table~\ref{tab:inputs} and the decay constant $f_{B_s}$ from Ref.~\cite{Tanabashi:2018oca}.

\subsubsection{FET vs. mass diagonalization}\label{sssec:FETcheck}

From the previous analyses, we can see that, in some regions of the NMSSM parameter space, the usually adopted MIA is not adequate by considering only the first one or two MI orders, and higher orders in the MI expansion must be considered. It is also shown that the FET results with optimal cutting-off MI orders do demonstrate good convergence. In this subsection, we show that these results also agree well with the ones calculated in the ME basis with exact diagonalization of the mass matrices that is achievable only numerically~\cite{Dedes:2015twa,Crivellin:2018mqz}.

\begin{figure}[htbp]
	\centering
	\includegraphics[width=0.9\textwidth]{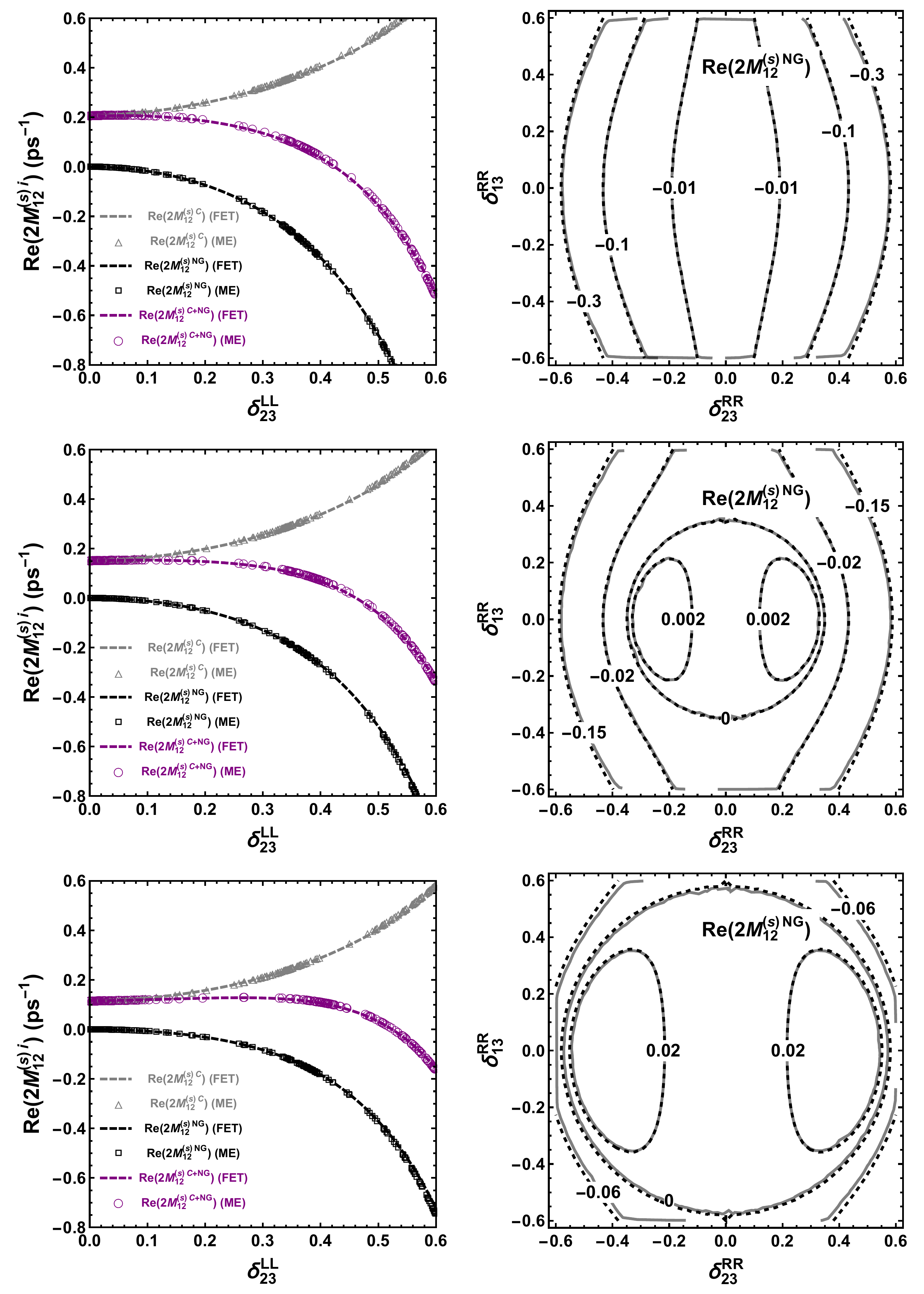}
	\caption{\label{FETMASS} \small The NMSSM contributions to $\text{Re}(2 M_{12}^{(s)\,i})$ in case~\textrm{IA} from different gauginos (left panel) and to $\text{Re}(2 M_{12}^{(s)\,\text{NG}})$ in case~\textrm{IIA} (right panel). Here $\delta_{33}^{\text{LR}}$ is fixed at $-0.15$ in case~\textrm{IA}, and $\sqrt{{M_S}_2}$ is set to be $1100$ (upper), $1300$ (middle), and $1500~\text{GeV}$ (lower), respectively. The dashed curves represent the FET results with squark MI order of $8$ in both cases, while the dotted with different geometries in the left and the solid in the right panel correspond to the ME calculations.}
\end{figure}

As an example, we show in Figure~\ref{FETMASS} the NMSSM contributions to $\text{Re}(2 M_{12}^{(s)\,i})$ from different gauginos~(see Eq.~\eqref{eq:m12decomposition} for their respective definitions) obtained with these two methods. Here only the NMSSM contribution to $\text{Re}(2 M_{12}^{(s)\,\text{NG}})$ is shown in case~\textrm{IIA}, because only the parameters $\delta_{13}^{\text{RR}}$ and $\delta_{23}^{\text{RR}}$ from $M_{\tilde{D}}^2$ are involved in this case, and they do not contribute to $\text{Re}(2 M_{12}^{(s)\,\text{C}})$. It can be seen clearly that the FET results with squark MI order of $8$ agree quite well with the ones calculated directly in the ME basis. One can also find from the left panel that, as the MI parameter $\delta_{23}^{\text{LL}}$ increases with the other parameters fixed in case~\textrm{IA}, the NMSSM contribution to $\text{Re}(2 M_{12}^{(s)\,\text{C}})$ becomes larger, while both $\text{Re}(2 M_{12}^{(s)\,\text{NG}})$ and $\text{Re}(2 M_{12}^{(s)\,\text{C}+\text{NG}})$ become smaller; in addition, their dependence on the parameter ${M_S}_2$ becomes weaker when $\sqrt{{M_S}_2}$ varies from $1100$~GeV to $1500$~GeV. In case~\textrm{IIA}, on the other hand, the variations of $\text{Re}(2 M_{12}^{(s)\,\text{NG}})$ with respect to  $\delta_{23}^{\text{RR}}$ and $\delta_{13}^{\text{RR}}$ depend heavily on the chosen values of $\sqrt{{M_S}_2}$: while the contours for negative $\text{Re}(2 M_{12}^{(s)\,\text{NG}})$ are similar, the trends for the extremum values of $\text{Re}(2 M_{12}^{(s)\,\text{NG}})$ with $\sqrt{{M_S}_2}=1100$~GeV are quite different from the ones with $\sqrt{{M_S}_2}=1300$~GeV or $1500$~GeV. All these observations can be more conveniently and easily understood in terms of the FET result for $\text{Re}(2 M_{12}^{(s)\,i})$, which is essentially a polynomial with the MI parameters as the variables.

\subsection{Constraints on the \boldmath{$\mathbb{Z}_3$}-invariant NMSSM parameters}\label{subsec:4.1}

As mentioned in the Introduction, the SM predictions for the mass differences $\Delta M_{s}$ and $\Delta M_{d}$ are now larger than their respective experimental values. For the time-integrated branching ratio $\overline{\mathcal B}(B_s \to \mu^+ \mu^-)$, on the other hand, the 2017 LHCb measurement\footnote{This decay has been searched for by CDF~\cite{Aaltonen:2013as} and D0~\cite{Abazov:2013wjb}, and was observed for the first time by LHCb~\cite{Aaij:2013aka} and CMS~\cite{Chatrchyan:2013bka}, with their combined average for the branching ratio given in Ref.~\cite{CMS:2014xfa}. Searches for this decay have also been performed by ATLAS~\cite{Aaboud:2016ire}. The 2017 LHCb measurement includes the LHC Run 2 data and represents the first single-experiment observation of this decay, with a $7.8\sigma$ significance~\cite{Aaij:2017vad}.}, $(3.0\pm0.6^{+0.3}_{-0.2})\times10^{-9}$~\cite{Aaij:2017vad}, is in reasonable agreement with the SM prediction, $(3.6 \pm 0.3)\times10^{-9}$. This will impose much more stringent constraints on NP~\cite{Altmannshofer:2017wqy}.

The inclusive radiative decay $B\to X_s\gamma$ provides also important constraints on NP scenarios with extended Higgs sectors or SUSY theories~\cite{Buras:1993xp,Hurth:2003vb,Hurth:2010tk,Paul:2016urs}. Measurements of its CP- and isospin-averaged branching ratio by the BaBar~\cite{Aubert:2007my,Lees:2012wg,Lees:2012ym} and Belle~\cite{Saito:2014das,Belle:2016ufb} collaborations lead to the following combined result~\cite{Misiak:2017bgg}
\begin{align}
{(\mathcal B_{s \gamma}^{\text{exp}})}_{E_{\gamma}>1.6\,\text{GeV}}=(3.27\pm0.14)\times10^{-4},
\end{align}
which is in excellent agreement with the state-of-the-art SM prediction~\cite{Misiak:2015xwa}
\begin{align}\label{eq:BsgammaSMMisiak}
{(\mathcal B_{s \gamma}^{\text{SM}})}_{E_{\gamma}>1.6\,\text{GeV}}=(3.36\pm0.23)\times10^{-4}.
\end{align}
Here the photon energy cutoff $E_{\gamma}>1.6\,\text{GeV}$ is defined in the decaying meson rest frame. The charged-Higgs contribution in the NMSSM (or MSSM) belongs to the case of Model-II considered in Ref.~\cite{Misiak:2017bgg}, and always interferes with the SM one in a constructive manner. Then, the extra one-loop SUSY contributions should involve a cancellation with that from the charged Higgs, so as to comply with the $B\to X_s\gamma$ constraint~\cite{Jager:2008fc,Buras:2002vd}. As the SM and charged-Higgs contributions to $\mathcal B_{d \gamma}$, the branching ratio of $B\to X_d\gamma$ decay, are both suppressed by the CKM factor $\left|K_{31}/K_{32}\right|^2$ with respect to $\mathcal B_{s \gamma}$, while the contributions from the squark MI parameters are not affected by this factor, it is expected that $\delta_{13}^{\text{LL}}$ in $M_{\tilde{U}}^2$ will be strongly constrained by $\mathcal B_{d \gamma}$. As a result, we have set $\delta_{13}^{\text{LL}}$ to be zero in Sec.~\ref{sec:FlavorStructure} from the very beginning. In this work, we use the public code {\tt SUSY\_FLAVOR}~\cite{Rosiek:2010ug,Crivellin:2012jv,Rosiek:2014sia} to get the NP contribution to $\mathcal B_{s \gamma}$ numerically, by adapting the initial values of input parameters to our case; especially, one must change the related parameters in the file to make sure that the photon energy cutoff $E_{\gamma}>1.6\,\text{GeV}$.

In the following, we shall exploit the $95\%$ C.L. bounds from $\Delta M_{s}$, $\Delta M_{d}$, $\overline{\mathcal B}(B_s \to \mu^+ \mu^-)$, and $\mathcal B_{s \gamma}$, to set the allowed regions for the NMSSM parameters ${M_S}_2$, $\delta_{23}^{\text{LL}}$, $\delta_{33}^{\text{LR}}$, $\delta_{13}^{\text{RR}}$, and $\delta_{23}^{\text{RR}}$, during which  only the experimental and the SM theoretical uncertainties are considered. Explicitly, we firstly construct an allowed range for each observable by taking into account both the experimental data and the SM prediction with their respective $1.96\sigma$ uncertainties added in quadrature, and then require the NMSSM central values to lie within the range, to get the allowed regions for the NMSSM parameters.  

\subsubsection{Results in scenario A}\label{sec:4.1.1}

\begin{figure}[htbp]
	\centering
	\includegraphics[width=0.45\textwidth]{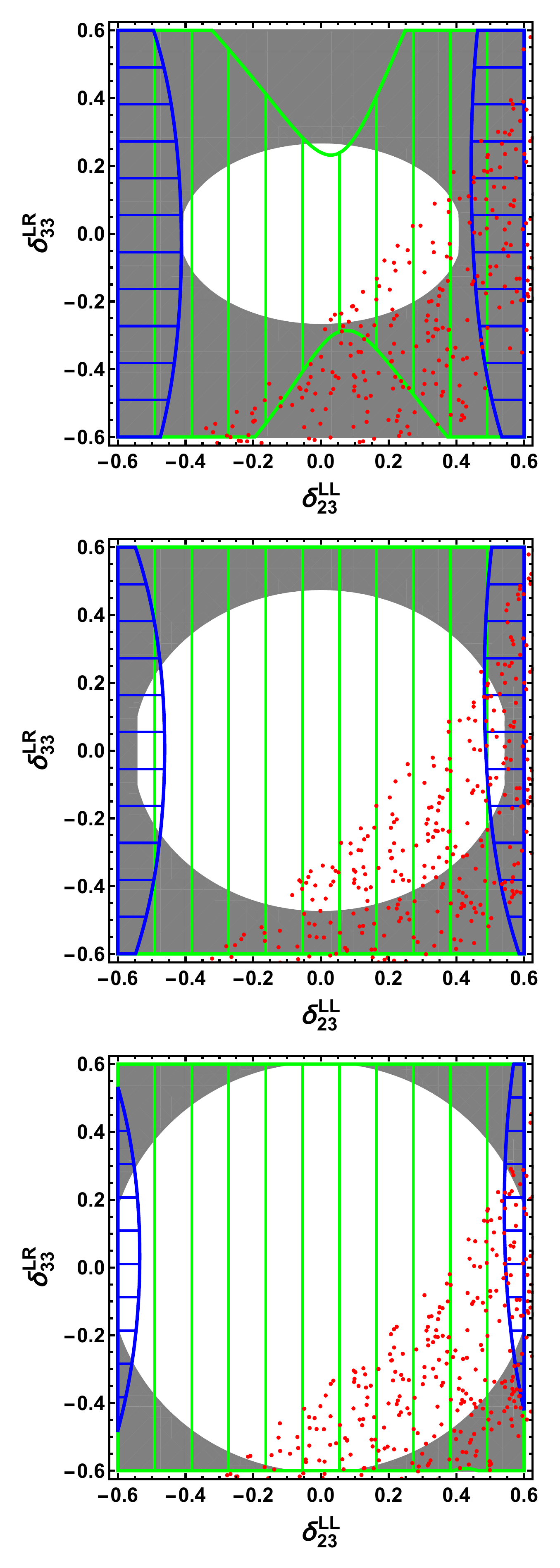}\hspace{0.2cm}
	\includegraphics[width=0.45\textwidth]{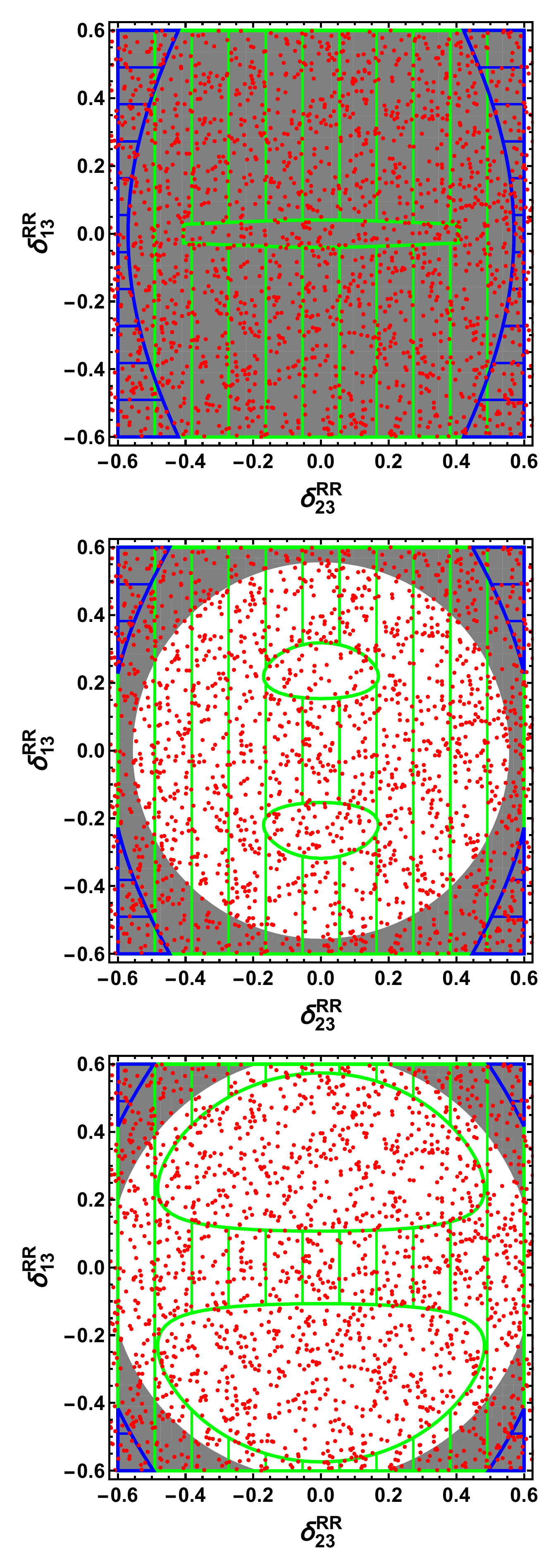}
	\caption{\label{fig:TP3REA} \small Allowed regions for $\delta_{23}^{\text{LL}}$, $\delta_{33}^{\text{LR}}$ in case \textrm{IA} (left panel) and $\delta_{13}^{\text{RR}}$, $\delta_{23}^{\text{RR}}$ in case \textrm{IIA} (right panel), respectively. Here $\delta_{33}^{\text{LR}}$ is fixed at $-0.4$ in cases \textrm{IIA} and $\sqrt{{M_S}_2}$ is set to be $1100$ (upper), $1300$ (middle), and $1500~\text{GeV}$ (lower), respectively. The red dot area is allowed by $\mathcal B_{s \gamma}$, while the green and the blue hatched area by $\Delta M_d$ and $\Delta M_s$, respectively. The gray areas are excluded by the lower bound of $940$~GeV on squark masses based on a rough estimate of the LHC direct searches for squarks~\cite{Sirunyan:2017leh,Aaboud:2017phn,Aad:2015gna,Sirunyan:2017xse,Sirunyan:2017pjw,Sirunyan:2017kiw,Aaboud:2017ayj,Aaboud:2017aeu}.}
\end{figure}

Firstly, we show in Figure~\ref{fig:TP3REA} the allowed regions for the squark MI parameters $\delta_{23}^{\text{LL}}$, $\delta_{33}^{\text{LR}}$, $\delta_{13}^{\text{RR}}$ and $\delta_{23}^{\text{RR}}$ in scenario A. Here, to match the SM-like Higgs mass, as shown in Figure~\ref{higgs125}, three choices of ${M_S}_2$ from $1100^2$ to $1500^2~\text{GeV}^2$ are made in both cases~\textrm{IA} and \textrm{IIA}, and the squark MI parameter $\delta_{33}^{\text{LR}}$ is set to be $-0.4$ in case~\textrm{IIA}. The scenario A is featured by the observation that the charged-Higgs contribution to $\Delta M_s$ is positive and large for small charged-Higgs mass and $\tan\beta$, while its contribution to $\mathcal B_{s \gamma}$, being also positive and large for small charged-Higgs mass, is less sensitive to the variation of $\tan \beta$. The bound from $\overline{\mathcal B}(B_s \to \mu^+ \mu^-)$ is, however, not shown in Figure~\ref{fig:TP3REA}, because on the one hand this bound is satisfied in the whole parameter regions displayed in Figure~\ref{fig:TP3REA} for case \textrm{I}, and on the other hand $\overline{\mathcal B}(B_s \to \mu^+ \mu^-)$ is nearly not affected by $\delta_{13}^{\text{RR}}$ or $\delta_{23}^{\text{RR}}$ in case \textrm{II}~\cite{Silvestrini:2007yf,Kumar:2016vhm}. One can see clearly that the excluded area by the lower bound of $940$~GeV on squark masses based on a rough estimate of the LHC direct searches for squarks~\cite{Sirunyan:2017leh,Aaboud:2017phn,Aad:2015gna,Sirunyan:2017xse,Sirunyan:2017pjw,Sirunyan:2017kiw,Aaboud:2017ayj,Aaboud:2017aeu} becomes reduced with increasing ${M_S}_2$.

The allowed region for $\delta_{23}^{\text{LL}}$ from $\Delta M_s$ also becomes reduced when ${M_S}_2$ increases, and only the one with large magnitudes of $\delta_{23}^{\text{LL}}$ exists in case~\textrm{IA}. It is particularly observed that, for ${M_S}_2=1100^2~\text{GeV}^2$, there exists no allowed region for $\delta_{23}^{\text{LL}}$ and $\delta_{33}^{\text{LR}}$ in this case. In case~\textrm{IIA}, on the other hand, the bounds from $\Delta M_s$ and $\Delta M_d$ become stronger for a larger ${M_S}_2$, and the one from $\Delta M_s$ is so strong that it is no longer compatible with that from the lower bound of $940$~GeV on squark masses. The bound from $\mathcal B_{s \gamma}$ also shows that a larger magnitude of $\delta_{23}^{\text{LL}}$ or $\delta_{33}^{\text{LR}}$ is required in case~\textrm{IA}, but almost the whole area of $\delta_{13}^{\text{RR}}$ and $\delta_{23}^{\text{RR}}$ is allowed in case~\textrm{IIA}. All these features can be explained by the following observations: in scenario~\textrm{A}, as mentioned before, there exist considerably large and positive contributions to $\Delta M_s$ and $\mathcal B_{s \gamma}$ from the charged Higgs; then a large magnitude of $\delta_{23}^{\text{LL}}$ or $\delta_{23}^{\text{RR}}$ ($\delta_{33}^{\text{LR}}$) is needed to provide large but negative contributions to $\Delta M_s$ ($\mathcal B_{s \gamma}$), so as to cancel the charged-Higgs effects~\cite{Jager:2008fc,Buras:2002vd}. Especially in case~\textrm{IIA}, the chosen value of $-0.4$ for $\delta_{33}^{\text{LR}}$ is already large enough to cancel the charged-Higgs effect on $\mathcal B_{s \gamma}$, and the flavour-violating contribution from the RR sector is small, making the whole area of $\delta_{13}^{\text{RR}}$ and $\delta_{23}^{\text{RR}}$ being allowed by this observable. Compared to the case from $\Delta M_s$, the allowed regions for the squark MI parameters from $\Delta M_d$ are relatively large, because the charged-Higgs contribution to $\Delta M_d$ is suppressed by the CKM factors in both cases \textrm{IA} and \textrm{IIA}, with some areas being not allowed in case~\textrm{IIA} due to the effect of $\delta_{13}^{\text{RR}}$.

After taking into account the $95\%$ C.L. bounds from $\Delta M_{s}$, $\Delta M_{d}$, $\mathcal B_{s \gamma}$, $\overline{\mathcal B}(B_s \to \mu^+ \mu^-)$, as well as the SM-like Higgs mass, we find numerically that the squark MI parameters $\delta_{23}^{\text{LL}}>0.45$ and $|\delta_{33}^{\text{LR}}|\sim 0.15$, and the allowed region is severely small in case~\textrm{IA}. In case~\textrm{IIA}, on the other hand, there exists no allowed region for $\delta_{13}^{\text{RR}}$ and $\delta_{23}^{\text{RR}}$.

\subsubsection{Results in scenario B}

\begin{figure}[htbp]
	\centering
	\includegraphics[width=0.46\textwidth]{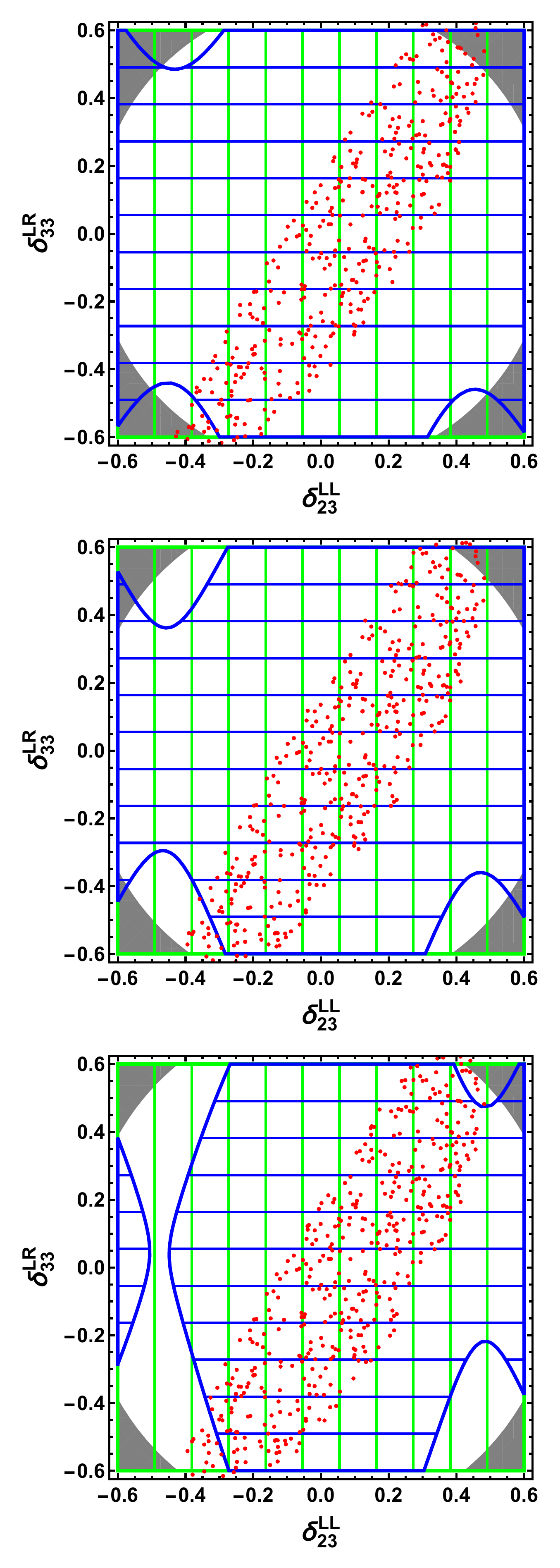}\hspace{0.2cm}
	\includegraphics[width=0.46\textwidth]{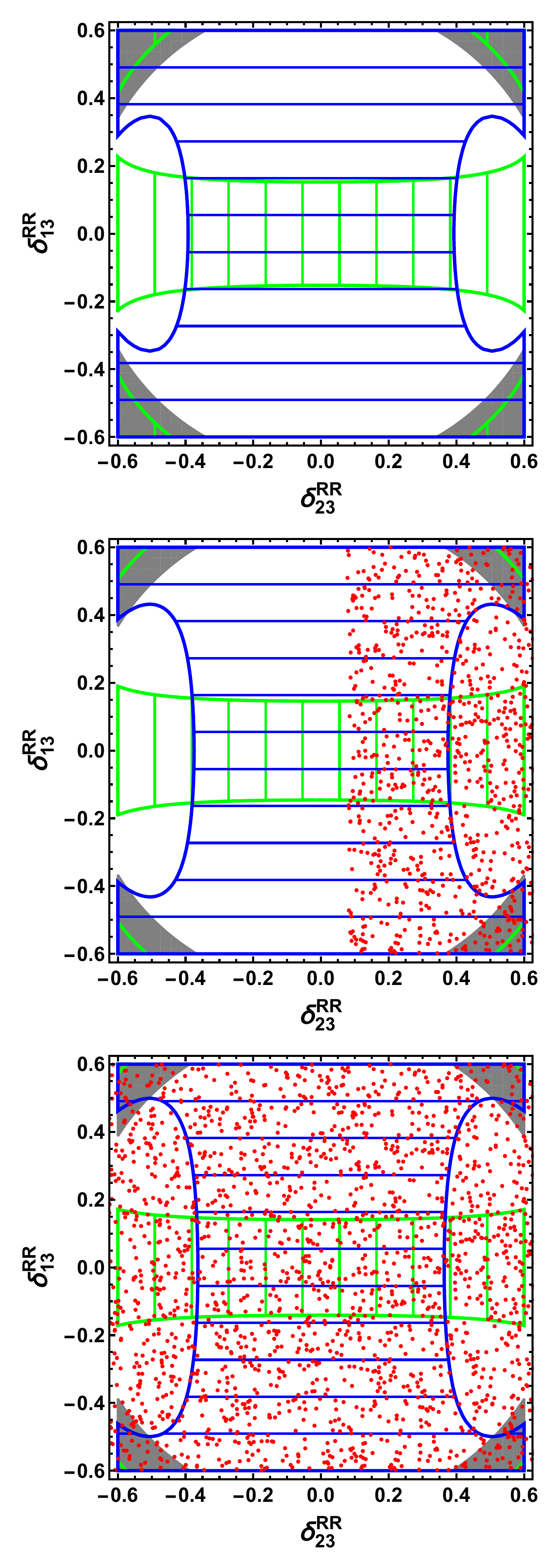}
	\caption{\label{fig:TP3REB} \small Allowed regions for $\delta_{23}^{\text{LL}}$ and $\delta_{33}^{\text{LR}}$ in case \textrm{IB} (left panel) as well as $\delta_{13}^{\text{RR}}$ and $\delta_{23}^{\text{RR}}$ in case \textrm{IIB} (right panel). Here $\delta_{33}^{\text{LR}}$ is fixed at $0.2$ in cases \textrm{IIB} and $\sqrt{{M_S}_2}$ is set to be $1700~(1800)$ (upper), $1800~(1900)$ (middle), and $1900~(2000)~\text{GeV}$ (lower) in case~\textrm{IB} (\textrm{IIB}). The other captions are the same as in Figure~\ref{fig:TP3REA}.}
\end{figure}

In Figure~\ref{fig:TP3REB}, we show the allowed regions for $\delta_{23}^{\text{LL}}$, $\delta_{33}^{\text{LR}}$, $\delta_{13}^{\text{RR}}$ and $\delta_{23}^{\text{RR}}$ in scenario B. To match the SM-like Higgs mass, as shown in Figure~\ref{higgs125}, the squark MI parameter $\delta_{33}^{\text{LR}}$ is set to be $0.2$ in case~\textrm{IIB}, and three choices of ${M_S}_2$ from $1700^2$~($1800^2$) to $1900^2$~($2000^2$)~$\text{GeV}^2$ are made in case~\textrm{IB} (\textrm{IIB}). The scenario B is featured by the observation that the positive contributions to $\Delta M_s$ and $\mathcal B_{s \gamma}$ with a large charged-Higgs mass are small, and considerably negative contributions from squarks and gauginos are not needed to cancel the charged-Higgs effects. As already observed in scenario A, the excluded area by the lower bound of $940$~GeV on squark masses also becomes reduced with increasing ${M_S}_2$, and the bound from $\overline{\mathcal B}(B_s \to \mu^+ \mu^-)$ is satisfied in the whole parameter regions shown here.

In case~\textrm{IB}, the allowed region for the squark MI parameters from $\Delta M_{s(d)}$ is large because of the small charged-Higgs contribution, and is almost independent of ${M_S}_2$; while the bound from $\mathcal B_{s \gamma}$ indicates that a smaller $\delta_{23}^{\text{LL}}$ is favored. In case~\textrm{IIB}, on the other hand, while the allowed region from $\Delta M_d$ stays nearly the same, the one from $\Delta M_s$ shrinks slowly when ${M_S}_2$ increases. Under the bound of $\mathcal B_{s \gamma}$, an allowed region with a positive $\delta_{23}^{\text{RR}}$ begins to emerge only when ${M_S}_2$ is about $1900^2~\text{GeV}^2$.

After considering the $95\%$ C.L. bounds from $\Delta M_{s}$, $\Delta M_{d}$, $\mathcal B_{s \gamma}$, as well as the SM-like Higgs mass, the parameter $\delta_{23}^{\text{LL}}$ can be small, and the allowed region is relatively larger in case~\textrm{IB} compared to the one in case~\textrm{IA}. In case~\textrm{IIB}, on the other hand, the allowed area of $\delta_{13}^{\text{RR}}$ and $\delta_{23}^{\text{RR}}$ exists only when ${M_S}_2$ is larger than about $1900^2~\text{GeV}^2$.

\section{Conclusion}\label{sec:conclusions}

In this paper, motivated by the observation that the SM predictions are now above the experimental data for the mass difference $\Delta M_{s(d)}$, and the usual CMFV models have difficulties in reconciling this discrepancy, we have investigated whether the $\mathbb{Z}_3$-invariant NMSSM with NMFV, in which the extra flavour violations arise from the non-diagonal parts of the squark mass squared matrices, can accommodate such a deviation, while complying with the experimental constraints from the branching ratios of $B_s\to \mu^+ \mu^-$ and $B\to X_s\gamma$ decays.

We have calculated the NMSSM contributions to $\Delta M_{s(d)}$ and the branching ratio $\overline{\mathcal B}(B_s \to \mu^+ \mu^-)$, using the recently developed FET procedure, which allows to perform a purely algebraic MI expansion of a transition amplitude written in the ME basis without performing the tedious and error-prone diagrammatic calculations in the interaction/flavour basis. Specifically, we have considered the finite MI orders for neutralinos but the general MI orders for squarks and charginos, under the following two sets of assumptions for the squark flavour structures: while the flavour-conserving off-diagonal element $\delta_{33}^\text{LR}$ is kept in both of these two sectors, only the flavour-violating off-diagonal elements $\delta_{23}^\text{LL}$ and $\delta_{i3}^\text{RR}$ ($i=1,2$) are kept in the \text{LL} and \text{RR} sectors, respectively. In this way, our analytic results are polynomials with the MI parameters as the variables and are expressed directly in terms of the initial Lagrangian parameters in the interaction/flavour basis, making it easy to impose the experimental bounds on them and, at the same time, allowing for a more transparent understanding of the qualitative behaviour of the transition amplitude. We have also presented an efficient method to estimate the optimal cutting-off MI orders for neutralinos, charginos, and squarks.

For the numerical analyses, we have considered two sets of NMSSM parameters that are denoted, respectively, by scenarios \textrm{A} and \textrm{B} in Table~\ref{tab:susyinputs}. Both of them can match the SM-like Higgs mass, and also make the Higgs-penguin effects negligible for $B_{s(d)}-\bar{B}_{s(d)}$ mixing and $B_s \to \mu^+ \mu^-$ decay. Together with the two assumptions for the squark flavour structures, there are totally four different cases, \textrm{IA}, \textrm{IB}, \textrm{IIA}, and \textrm{IIB}, to be discussed. Firstly, after getting the optimal cutting-off MI orders for neutralinos, charginos, and squarks with our estimation rules, we have verified the convergence of the FET results obtained with the corresponding MI orders, even in the case when the MI parameters are large. Then, taking $\text{Re}(2 M_{12}^{s})$ in cases \textrm{IA} and \textrm{IIA} as an example, we have demonstrated that the FET results with optimal cutting-off MI orders agree quite well with the ones calculated directly in the ME basis with an exact diagonalization of the mass matrices that is usually achievable only numerically. This also proves the necessity to consider the optimal cutting-off MI orders when performing an NMFV expansion, as the expansion at leading order is usually insufficient and could easily be misleading.

Finally, after considering the $95\%$ C.L. bounds from the observables $\Delta M_{s}$, $\Delta M_{d}$, $\overline{\mathcal B}(B_s \to \mu^+ \mu^-)$, $\mathcal B_{s \gamma}$, as well as the SM-like Higgs mass, we have discussed the allowed regions for the parameters ${M_S}_2$, $\delta_{23}^{\text{LL}}$, $\delta_{33}^{\text{LR}}$, $\delta_{13}^{\text{RR}}$, and $\delta_{23}^{\text{RR}}$. It is found that only large values of $\delta_{23}^{\text{LL}}$, with $|\delta_{33}^{\text{LR}}|\sim 0.15$, are allowed in case~\textrm{IA}, and the allowed region for $\delta_{23}^{\text{LL}}$ becomes reduced when ${M_S}_2$ increases from $1100^2$ to $1500^2~\text{GeV}^2$. In case~\textrm{IB}, on the other hand, with $\delta_{33}^{\text{LR}}$ being fixed at about $0.2$, relatively smaller magnitude of $\delta_{23}^{\text{LL}}$ is found to be allowed, and the allowed region for $\delta_{23}^{\text{LL}}$ becomes almost independent of ${M_S}_2$. In case~\textrm{IIA}, with $\delta_{33}^{\text{LR}}$ being fixed at $-0.4$, there exists no allowed region for $\delta_{13}^{\text{RR}}$ and $\delta_{23}^{\text{RR}}$, because of the strong bound from $\Delta M_s$. In case~\textrm{IIB}, on the contrary, with $\delta_{33}^{\text{LR}}$ being fixed at about $0.2$, the allowed region for $\delta_{13}^{\text{RR}}$ and $\delta_{23}^{\text{RR}}$ exists only when ${M_S}_2$ is larger than about $1900^2~\text{GeV}^2$.

As a final remark, we should mention that, with the experimental progress in direct searches for SUSY particles as well as the more and more precise theoretical predictions for these observables, the NMSSM effects on these low-energy flavour processes can be further exploited.

\section*{Acknowledgements}
We would like to thank Michael Paraskevas, Janusz Rosiek, and Xing-Bo Yuan for valuable discussions and be grateful to Michael Paraskevas for his guidance on {\tt SUSY\_FLAVOR}. This work is supported by the National Natural Science Foundation of China  under Grant Nos.~11675061, 11775092, 11521064, and 11435003. X.L. is also supported in part by the self-determined research funds of CCNU from the colleges' basic research and operation of MOE~(CCNU18TS029). Q.H. is also supported by the China Postdoctoral Science Foundation~(2018M632896).

\appendix
\renewcommand{\theequation}{A.\arabic{equation}}
\section*{Appendix: Block terms for squarks and charginos}
\label{app:block}

In this appendix, we list all the non-zero block terms for squarks and charginos. For convenience, we introduce the notations $\Delta_{1} \equiv \frac{{M_S}_1 {M_S}_2 \delta^2_{23}}{q^2-{M_S}_1}+\frac{{M^2_S}_2 \delta^2_{36}}{q^2-{M_S}_2}$ and
$\Delta^{\prime}_{1} \equiv \frac{{M_S}_1 {M_S}_2 (1+\lambda^2_{\text{CKM}}) \delta^2_{23}}{q^2-{M_S}_1}$ in case I, and $\Delta_{2} \equiv \frac{{M_S}_1 {M_S}_2 (\delta^2_{46}+\delta^2_{56}) }{q^2-{M_S}_1}$ in case II. In the following, $n=1, 2, 3, \cdots$, denotes the MI-order index.

For up-type squarks, the non-zero block terms are given as
\begin{align}
L_\text{U}(0;i,i)&=
\begin{cases}
\frac{1}{q^2-{M_S}_1},& i=1,2,4,5\\
\frac{1}{q^2-{M_S}_2},& i=3,6
\end{cases},\\[2mm]
L_\text{U}(2n-2;3,3)&=\frac{1}{(q^2-{M_S}_2)^{n}} \Delta^{n-1}_1 ,\\[2mm]
L_\text{U}(2n-1;3,i)&=
\begin{cases}
\frac{\sqrt{{M_S}_1 {M_S}_2}\delta_{23}}{(q^2-{M_S}_1)(q^2-{M_S}_2)^{n}} \Delta^{n-1}_1,& i=2\\
\frac{{M_S}_2 \delta_{36}}{(q^2-{M_S}_2)^{n+1}} \Delta^{n-1}_1,& i=6
\end{cases} ,\\[2mm]
L_\text{U}(2n;i,j)&=
\begin{cases}
\frac{{M_S}_1 {M_S}_2 \delta^2_{23}}{(q^2-{M_S}_1)^2 (q^2-{M_S}_2)^{n}}
\Delta^{n-1}_1, & (i,j)=(2,2)\\
\frac{{M_S}_2 \sqrt{{M_S}_1 {M_S}_2} \delta_{23} \delta_{36}}{(q^2-{M_S}_1) (q^2-{M_S}_2)^{n+1}}
\Delta^{n-1}_1, & (i,j)=(2,6),(6,2)\\
\frac{{M^2_S}_2 \delta^2_{36} }{(q^2-{M_S}_2)^{n+2}}
\Delta^{n-1}_1, & (i,j)=(6,6)
\end{cases} ,\\[2mm]
L_{\text{UU}}(2n;3,i;3,j)&=\sum_{n_1=1}^{n} L_\text{U}(2n_1-1;3,i) L_\text{U}(2(n+1-n_1)-1;3,j), \quad \text{$i,j=2$ or $6$,} \label{eq:LUU3i3j}\\
L_{\text{UU}}(2n-1;i,j;3,j')&=\sum_{n_1=0}^{n-1} L_\text{U}(2n_1;i,j) L_\text{U}(2(n-n_1)-1;3,j'), \quad \text{$i,j=1$--$6$ and $j'=2$ or $6$},\\
L_{\text{UU}}(2n-2;i,j;i',j')&=\sum_{n_1=0}^{n-1} L_\text{U}(2n_1;i,j) L_\text{U}(2(n-1-n_1);i',j') , \quad \text{$i,j,i',j'=1$--$6$},
\end{align}
in case I.

For down-type squarks, the non-zero block terms are given as,
\begin{align}
L_\text{D}(2n-1;3,i)&=
\begin{cases}
\frac{- \sqrt{{M_S}_1 {M_S}_2} \lambda_{\text{CKM}} \delta_{23} }{(q^2-{M_S}_1)(q^2-{M_S}_2)^{n}} {\Delta^{'}}^{n-1}_1,
& i=1\\[2mm]
\frac{  \sqrt{{M_S}_1 {M_S}_2}  \delta_{23} }{(q^2-{M_S}_1)(q^2-{M_S}_2)^{n}} {\Delta^{'}}^{n-1}_1,
& i=2
\end{cases} ,\\[2mm]
L_{\text{DD}}(2n;3,i;3,i)&=\sum_{n_1=1}^{n} L_\text{D}(2n_1-1;3,i) L_\text{D}(2(n+1-n_1)-1;3,i) , \quad \text{$i=1$ or $2$},
\end{align}
in case I, and
\begin{align}
L_\text{D}(2n-1;6,i)&=
\frac{\sqrt{{M_S}_1 {M_S}_2}\delta_{i6}}{(q^2-{M_S}_1)(q^2-{M_S}_2)^{n}} \Delta^{n-1}_2 ,\quad \text{$i=4$ or $5$},\\[2mm]
L_{\text{DD}}(2n;6,i;6,i)&=\sum_{n_1=1}^{n} L_\text{D}(2n_1-1;6,i) L_\text{D}(2(n+1-n_1)-1;6,i) ,\quad \text{$i=4$ or $5$},
\end{align}
in case II.

The non-zero block terms for charginos include
\begin{align}
L_{\text{X}}(2n-2;i,i)=\frac{1}{(q^2-{{M}_X}_i)^{n}(q^2-{{M}_X}_{i'})^{n-1}}  (\delta^{X}_{12} \sqrt{{{M}_X}_1 {{M}_X}_2})^{2n-2} ,
\end{align}
where $(i,i')=(1,2)$ or $(2,1)$,
\begin{align}
L_{\text{X}}(2n-1;1,2)=\frac{1}{(q^2-{{M}_X}_1)^{n}(q^2-{{M}_X}_2)^{n}}   (\delta^{X}_{12} \sqrt{{{M}_X}_1 {{M}_X}_2})^{2n-1} ,
\end{align}
\begin{align}
L_{\text{XY}}(2n-2;i,i;j,j)=\sum_{n_1=0}^{n-1} L_\text{X}(2n_1;i,i) L_\text{Y}(2(n-1-n_1);j,j),
\end{align}
where $(i,j)=(1,1),(1,2),(2,1)$, or $(2,2)$,
\begin{align}
L_{\text{XY}}(2n-1;i,i;1,2)=\sum_{n_1=0}^{n-1} L_\text{X}(2n_1;i,i) L_\text{Y}(2(n-n_1)-1;1,2),
\end{align}
where $i=1$ or $2$, and
\begin{align}
L_{\text{XY}}(2n;1,2;1,2)=\sum_{n_1=1}^{n} L_\text{X}(2n_1-1;1,2) L_\text{Y}(2(n+1-n_1)-1;1,2).
\end{align}
Here the subscripts $\text{X}$ and $\text{Y}$ can be $\text{C}$ or $\text{P}$.

\bibliographystyle{JHEP}
\bibliography{ref}

\end{document}